\documentclass[useAMS,usenatbib]{mn2e}
\usepackage{graphicx}
\usepackage{txfonts}
\usepackage{color,epsfig,amssymb,longtable}
\usepackage{array,colortbl,lscape}

\newcommand{\aap}{A\&A}

\newcommand{\halpha}{H$_{\alpha}$}
\newcommand{\hbeta}{H$_{\beta}$}

\newcommand{\Msun}{M$_{\odot}$}
\newcommand{\Lsun}{L$_{\odot}$}
\newcommand{\Zsun}{Z$_{\odot}$}
\bibpunct[,]{(}{)}{;}{a}{,}{,}

\begin{document}

\title[PopStar I: Evolutionary synthesis models]{PopStar I: Evolutionary
synthesis models description}

\author[Moll{\'{a}}, Garc{\'{\i}}a-Vargas and Bressan]{M.~Moll{\'{a}},$^{1}$,
 M.L. Garc{\'{\i}}a-Vargas,$^{2}$ and A. Bressan $^{3,4,5}$  \\
$^{1}$ Departamento de Investigaci\'{o}n B\'{a}sica, CIEMAT,
Avda. Complutense 22, 28040, Madrid, (Spain) \\
$^{2}$ FRACTAL SLNE, C/ Tulip\'{a}n 2, p13, 1A , 28231 Las Rozas de Madrid,
(Spain)\\
$^{3}$ INAF Osservatorio Astronomico di Padova, Viccolo dell' Osservatorio-5,
35122 Padova (Italy)\\
$^{4}$SISSA-ISAS, International School for Advanced Studies, Via Beirut 4, 34014 Trieste (Italy)\\
$^{5}$INAOE, Luis Enrique Erro 1, 72840, Tonantzintla, Puebla (Mexico)\\
}

\date{Accepted Received ; in original form }

\pagerange{\pageref{firstpage}--\pageref{lastpage}} \pubyear{2008}

\maketitle \label{firstpage}

\begin{abstract}
We present new evolutionary synthesis models for Simple Stellar
Populations for a wide range of ages and metallicities. The models are
based on the Padova isochrones.  The core of the spectral library is
provided by the medium resolution Lejeune et al. atmosphere models.
These spectra are complemented by NLTE atmosphere models for hot stars
that have an important impact in the stellar cluster's ionizing
spectra: O, B and WR stellar spectra at the early ages, and spectra of
post-AGB stars and planetary nebulae, at intermediate and old ages.
At young ages, our models compare well with other existing models but
we find that, the inclusion of the nebular continuum, not considered
in several other models, reddens significantly the integrated colours
of very young stellar populations. This is consistent with the results
of spectral synthesis codes particularly devised for the study of
starburst galaxies.  At intermediate and old ages, the agreement with
literature model is good and, in particular, we reproduce well the
observed colours of star clusters in LMC.  Given the ability to
produce good integrated spectra from the far-UV to the infrared at any
age, we consider that our models are particularly suited for the study
of high redshift galaxies.  These models are available on the web site
{http://www.fractal-es.com/SEDmod.htm} and also through the Virtual
Observatory Tools on the PopStar server.
\end{abstract}

\begin{keywords} galaxies: abundances -- galaxies: evolution--  galaxies:
starburst --galaxies: stellar content \end{keywords}

\section{Introduction}

Spectro-photometric evolutionary synthesis models are the basic tools
to obtain information about age, metallicity, mass, star formation
history and other properties of objects whose stellar populations are
unresolved.  Population synthesis has proven to be now even more
important to analyze data obtained from large surveys or/and from high
redshift galaxies and to extract the star formation histories of
millions of galaxies \citep*[e.g.][]{cle06,asa07,cid07,ann07,cle09a}.

Almost all the models used to study the integrated properties of
stellar systems rest now on the notion, first introduced by Tinsley
(1972), that the stellar birthrate can be split into the product of a
time independent function providing the star mass distribution, the
initial mass function (IMF), and a mass independent function providing
the star formation rate.  In this way the integrated properties of a
galaxy can be modeled as a combination of Simple Stellar Populations
(SSP), that can be considered the building blocks of the population
synthesis technique \citep{bru83}.

There have been in the past many studies devoted to the computation of
integrated properties of simple stellar populations, e.g.
\citet*{bre94,bc03,stb99,bick04,gon05,mar05,fri06}.  Important
differences between these SSP models arise from the use of different
stellar tracks (or isochrones), different stellar atmosphere
libraries, different spectral coverage and resolution, inclusion of
nebular emission and different input physics and computational
algorithms.  It is also worth noticing that many of the existing codes
are optimized for the application to particular types of objects. For
example, \citet[][ hereafter STB99]{stb99} is tuned to analyze
starburst galaxies, while other standard evolutionary synthesis models
are mainly suited for old stellar populations.

Particularly important today, where a large observational effort is
devoted to high redshift galaxies, is the ability to accurately
describe the concomitant presence of very young and intermediate age
(from a fraction to a few Gyr) stellar populations.  Indeed, in the
studies of high redshift galaxies, it has been shown that the use of
models based on different ingredients may change dramatically the
estimates of ages and masses of the underlying stellar populations
\citep{mar05, bru07}.

This piece of work is the first paper of a series of three dedicated
to PopStar models description and initial test-cases application.  In
this paper, Paper I, we describe PopStar models and show their good
calibration with previous ones. We emphasize their suitability to
model stellar populations in a wide range of age and
metallicity. Paper II summarizes the results of computing the emission
line spectra of the HII regions ionized by the youngest star clusters
and compares these results with a sample of HII regions where the
metallicity has been carefully determined through appropriate
calibrators. Finally, Paper III calculates the photometrical
properties of young star clusters, taking into account the
contamination by the emission lines in the computed colors and the
underlying old populations where these clusters are embedded with the
aim of being able to determine the cluster physical properties from
the photometrical information only in order to apply to large
photometrical surveys.

The aim of this paper is to present an updated version of our
evolutionary synthesis code \citep*{gmb98,mgv00} that makes it suitable
for the study of a wide range of stellar populations, from those
extremely young to the oldest one, in a widest as possible spectral
range.  This is required on one side by the increasing relevance of
young and intermediate age stellar populations in observed magnitudes
and colours of high redshift galaxies and, on the other, by the advent
of multi-band observations that cover at least from the rest frame far
UV to beyond the near infrared.

We present a new code that combines a revision of the Padova
isochrones used in \citet{gmb98}, mainly in the computation of
intermediate age stellar populations, with the best updated stellar
model spectra, including nebular emission for the youngest ages.  The
SSP spectra are based on the stellar atmosphere models by \citet*{lcb97}
and \citet*{snc02} for normal and massive stars, respectively, and the
planetary nebula models by \citet{rau03}.  The main reason for avoiding
the use of empirical libraries, is that they still provide an
incomplete coverage of the parameter space (effective temperature,
gravity, metallicity, wavelength).  The reason to avoid the use of
other theoretical atmosphere libraries with good parameter space
coverage and even finer spectral resolution, is that the latter models
have not been yet carefully tested against observations as the one we
have adopted.  The models presented here are thus at medium spectral
resolution and they have not been computed for comparison with high
spectral resolution observations, like those recently published by
\citet{gon05,coe07}.  However there are not, at the moment, high
resolution spectra for all the stellar phases considered in this
paper, and thus we paid more attention to the wide wavelength
coverage, because our goal is to provide models useful for the study
of high redshift galaxies.

Finally we remind that in the last decade it was recognized the
importance of considering the effects of dust, not only for the
extinction but also for the re-emission of the light in the mid and
far infrared spectral regions, with the consequent extension of the
population synthesis into these new spectral windows
\citep*{bgs98,bre06,mar08,cle09b} and even into the radio window
\citep{bre02}.  As for the majority of literature models, also those
presented here do not account for this effect directly, but they can
be used in existing galaxy spectral evolution codes that account for
the effects of dust as e.g. GRASIL \citep*{sil98, pan03,veg08}.

The plan of the paper is the following.  In Section 2 we summarize the
main characteristics of evolutionary tracks (Padova 94), isochrone
calculation and stellar atmosphere models adopted.  In Section 3 we
illustrate and discuss the inclusion of the nebular contribution. In
Section 4 we present the photometrical evolution of our SSP and
compare it with that obtained by other selected models existing in
literature.  Finally, our conclusions are drawn in Section 5.

\section{The Evolutionary Synthesis Model}
\subsection{The stellar evolution: the isochrones}

The grid is composed by Simple Stellar Populations (SSP) for six
different IMF's whose characteristics are summarized in
Table~\ref{imfs}.  The \cite{sal55} power law, $\phi(m) \propto
m^{-(\alpha+1)} $ with $\alpha = $1.35 with mass limits: a)
between 0.85 and 120 M$_{\odot}$, and b) between 1.00 and 100
M$_{\odot}$, have been used for comparison with other set of
models, in particular \citep*{gbd95a} and STB99 \citep{stb99},
respectively.  The others IMF functions listed in Table~\ref{imfs} are
\cite{sal55}, \citet*{fer90},\citet{kro01} and \citet*{cha03}, all of them
computed with masses between 0.15 and 100 M$_{\odot}$. The expressions
are:

\begin{eqnarray}
\phi(m)_{SAL} & = &  m^{-2.35} \mbox{} \\
\phi(m)_{FER} & = & 10^{-\sqrt{0.73+\log{m} (1.92+\log{m} 2.07)}}/m^{1.52}
\mbox{ } \\
\phi(m)_{KRO} & = & \left\{ \begin{array}{l r}
  m^{-0.35}    &  \mbox{ $0.15 \le m/M_{\odot} < 0.08$} \\
 0.08 m^{-1.3} &  \mbox{ $0.08 \le m/M_{\odot} < 0.50$} \\
 0.04 m^{-2.3} &  \mbox{ $0.50 \le m/M_{\odot} < 100$ } \end{array} \right. \\
\phi(m)_{CHA} & = & \left\{ \begin{array}{l r}
0.037 m^{-1} e^{-\frac{(\log{m}+0.658)^{2}}{0.65}}& \mbox{$0.15 \le m/M_{\odot} < 1$} \\
0.019 m^{-2.3} & \mbox{$1 \le m/M_{\odot} \le 100$} \end{array} \right.
\label{imfs}
\end{eqnarray}

The isochrones used in this work have been computed for six different
metallicities: Z $ = $ 0.0001, 0.0004, 0.004, 0.008, 0.02 and 0.05
adopting stellar evolutionary tracks from \cite{bre93,fagotto94a,fagotto94b} 
and \cite*{girardi96}.  While the general properties of
the stellar evolutionary tracks can be found in the above papers, we
stress here a few points that are relevant for the present paper.

{\it a) Mass loss}
Mass-loss is an important ingredient in the context of this paper
because, in old and intermediate age stellar populations it determines
the current stellar mass on the Horizontal Branch and on the Post
Asymptotic Giant Branch (P-AGB) phases while, in young populations, it
affects the relative distribution between supergiant and Wolf Rayet
single stars.

Along the red giant branch (RGB) phase of low mass stars, mass loss is
accounted for by means of the Reimers formulation, assuming
$\eta_{RGB}$ = 0.50 independently of the metallicity as suggested by
\citet*{carraro96} and more recently by \citet*{vanloon08}.  The AGB
phase of old and intermediate age populations is calculated according
to \citet{bgs98}.  The adopted value of $\eta_{RGB}$ is slightly larger
than that assumed by \cite{bgs98}, $\eta_{RGB}$ = 0.45.  However with
these prescriptions the isochrones reproduce fairly well the UV
integrated properties of Globular Clusters \citep{Chavez09}, and the
observed Spitzer IRS spectra and mid infrared colours of ellipticals
in Virgo \citep{bre06}, and Coma \citep{cle09b}, clusters,
 i.e. old stellar populations likely at the two extremes
of the metallicity range of stellar systems.  At intermediate ages,
the colours compare well with the observations of LMC clusters
\citep*{per83,kye03,goud06,pes06} as shown later in Figure \ref{vktest}.

The mass loss formulation adopted in massive stars is amply described
in the relative quoted papers.  Here we remind that it was included in
the computation of the evolutionary tracks with a metallicity
dependence $\sqrt(Z)$, {\it independently from the evolutionary
status}.  The surface hydrogen, helium and carbon abundances are used
to discriminate blue supergiant form Wolf Rayet (WR) stars, and to
identify the main WN and WC evolutionary phases, as described in
\citet{gbd95a}.  In this respect we remind that, besides mass loss, the
appearance of the WR phase depends strongly also on other input
physics like the stellar rotation \citep{meynet06, vazquez07} and/or
the evolution within a binary system \citep[e.g][]{van07} with
suitable orbital parameters. These effects are not included in our
treatment.

{\it b Ages}
The age coverage is from $\log{\tau}=$ 5.00 to 10.30 with a variable
time resolution which is $\Delta(\log{\tau})=0.01$ in the youngest
stellar ages.  At ages of a few million years, low mass stars are
still contracting on the main sequence.  This is not very important
for the integrated properties because, under our IMF assumptions, the
integrated luminosity is dominated by the most massive stars. At these
young ages however, the latter stars may be still embedded in their
parent molecular cloud, with the consequence that their ionizing
spectra may be heavily modified by extinction.  The effect in these
cases would be that dust may attenuate the flux of ionizing photons,
changing the total intensity and hardness of the ionizing spectrum.
This effect is not considered in this paper but it will be the subject
of a forthcoming investigation devoted to the effects of dust in young
star forming regions. Nevertheless, our results are applicable to many
star formation regions (e.g. Giant Extragalactic HII regions) that are
not dominated by dust (massive stars are cleary visible even at UV
images taken with HST).

{\it c) Stochastic effects}

When dealing with relatively low populated stellar systems it has been
suggested that stochastic effects may significantly affect the
comparison between models and observations.  This is a wavelength
dependent effect because stars in different evolutionary phases do not
affect the various spectral regions in the same way.  Several papers
have been devoted to the analysis of stochastic effects in
evolutionary synthesis models
\citep*[e.g.][]{chio88,ceal02,cl04,cl06,clj08}.  Tuning the content of
massive stars in star-forming regions via mass segregation related to
stochastic effect was suggested by \citet{gvd94} as an alternative to
an IMF steepening to decrease the number of massive stars, what helped
to explain the emission line ratios observed in Giant Extragalactic
high metallicity H{\sc ii} regions. As we will show in Paper II these
observations can now naturally be fitted by PopStar models, without
the need of introducing this effect. Moreover, observations with HST
show evidence against these stochastic effects in star forming
regions, since very massive stars have been detected even in small
clusters \citep[e.g.][]{wmb02,umm07}.

\begin{table}
\caption{Summary of the used IMF's}
\begin{tabular}{cccc}
\hline
IMF & $\rm M_{low}$ &  $\rm M_{up}$ &  Reference \\
    & (M$_{\odot}$) &  (M$_{\odot}$)&  \\
\hline
SAL1  & 0.85 & 120 & Salpeter (1955)\\
SAL2  & 0.15 & 100 & Salpeter (1955)\\
FER   & 0.15 & 100 & Ferrini et al. (1989)\\
CHA   & 0.15 & 100 & Chabrier (2003)\\
KRO   & 0.15 & 100 & Kroupa (2001) \\
STB   & 1.00 & 100 & Salpeter (1955)\\
\hline
\end{tabular}
\label{imfs}
\end{table}

\begin{figure}
\resizebox{\hsize}{!}{\includegraphics[angle=0]{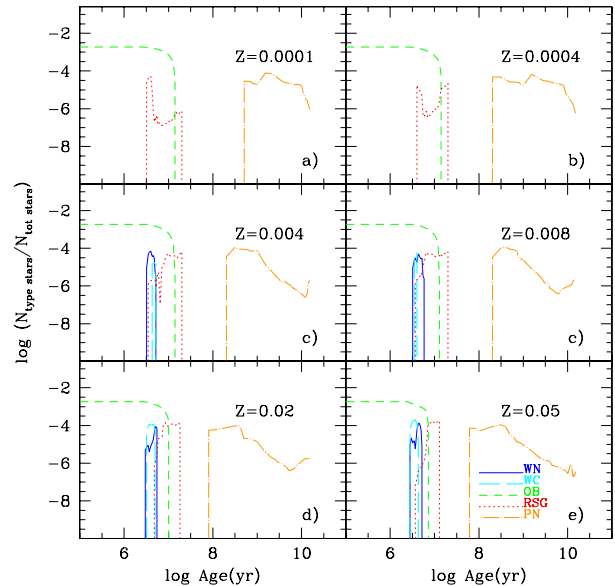}}
\caption{Logarithm of the ratio between different type of stars and the
total number of stars along 15 Gyr evolution. SSPs computed for IMF by
Salpeter with $\rm m_{low}=0.15 M_{\odot}$, $\rm m_{up}= 100
M_{\odot}$ and 6 metallicities, Z: a) 0.0001, b) 0.0004, c) 0.004, d)
0.008, e) 0.02 and f) 0.05}
\label{nest_z}
\end{figure}

\begin{figure}
\resizebox{\hsize}{!}{\includegraphics[angle=-90]{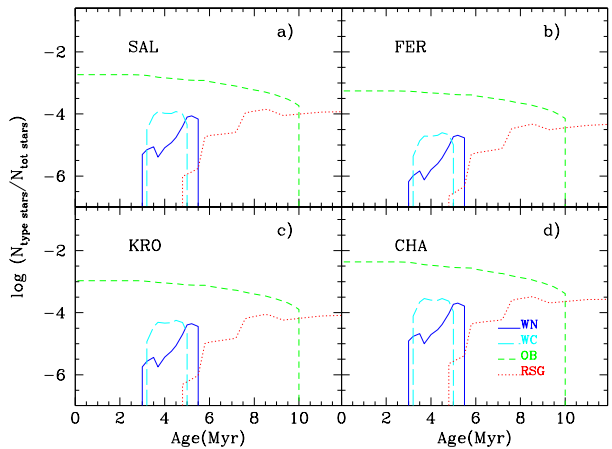}}
\caption{Logarithm of ratio between different type of stars and the
total number of stars, along the first 5 Myr of the cluster evolution
for solar metallicity (Z = 0.02) and 4 different IMF's: a) SAL2, b) FER,
c) KRO and d) CHA.  All of them defined in the same mass range,
between $\rm m_{low}=0.15 M_{\odot}$ and $\rm m_{up}=100 M_{\odot}$.}
\label{nest_imf}
\end{figure}

We have defined stellar groups along each isochrones:
Massive stars, MAS, are all stars with initial mass larger than 20
M$_{\odot}$; OB stars when their effective temperature is Teff $\ge$
25000~K; red supergiants (RSG) if Teff $<$ 6760~K; Wolf Rayet stars
are split into WN stars, when the surface hydrogen content by mass is
X$<$0.3, and WC stars when at the surface X = 0 and $\rm ^{12}C >
^{14}$N by number.  All stars more massive than 8 M$_{\odot}$ are
supposed to die as core collapsed supernovae (SN).  Intermediate and
low mass (M $< 15 M_{\odot}$ stars in the evolutionary phases beyond
the asymptotic giant branch phase with Teff$>$ 50000~K, are named
Planetary Nebula (PN) stars.  All other stars are classified as {\it
normal stars}, NS.  This classification will be useful to assign
a suitable spectrum to the corresponding star.

Table~\ref{contributions} shows, as an example, the time evolution of
the number of massive stars, normalized to an initial cluster mass of
1 M$_\odot$, for a Salpeter IMF with m$_{low} = $ 0.15 and m$_{up} = $ 100
(SAL2) and for solar metallicity.
\begin{table*}
\caption{Number of stars of different type in each age. We list only
the results for ages up to 5 Myr. We consider in this table numbers
normalized to a cluster mass of 1 M$_{\odot}$, a Salpeter IMF with
mlow$ = 0.15 M_{\odot}$, mup$ =  100
M_{\odot}$, and Z$ = 0.02$.  The table for the whole range of age and
metallicity is available in electronic format.}
\begin{tabular}{cccccccc}
\hline
 $\log{Age}$(yr) & N$({\rm M>20M_{\odot}}$) &  N$_{\rm OB}$  &
N$_{\rm WR-WN}$ & N$_{\rm WR-WC}$ & N$_{\rm RSG}$ &  N$_{\rm SNII/Ib/Ic}$&
N$_{\rm PN}$  \\
\hline
 5.00 &2.311e-03& 3.555e-03 &0.000e+00& 0.000e+00 &0.000e+00& 0.000e+00 &0.000e+00  \\
 5.48 &2.311e-03& 3.555e-03 &0.000e+00& 0.000e+00 &0.000e+00& 0.000e+00 &0.000e+00  \\
 5.70 &2.311e-03& 3.555e-03 &0.000e+00& 0.000e+00 &0.000e+00& 0.000e+00 &0.000e+00  \\
 5.85 &2.311e-03& 3.555e-03 &0.000e+00& 0.000e+00 &0.000e+00& 0.000e+00 &0.000e+00  \\
 6.00 &2.311e-03& 3.555e-03 &0.000e+00& 0.000e+00 &0.000e+00& 0.000e+00 &0.000e+00  \\
 6.11 &2.312e-03& 3.555e-03 &0.000e+00& 0.000e+00 &0.000e+00& 0.000e+00 &0.000e+00  \\
 6.18 &2.311e-03& 3.555e-03 &0.000e+00& 0.000e+00 &0.000e+00& 0.000e+00 &0.000e+00  \\
 6.23 &2.312e-03& 3.555e-03 &0.000e+00& 0.000e+00 &0.000e+00& 0.000e+00 &0.000e+00  \\
 6.30 &2.311e-03& 3.555e-03 &0.000e+00& 0.000e+00 &0.000e+00& 0.000e+00 &0.000e+00  \\
 6.34 &2.312e-03& 3.555e-03 &0.000e+00& 0.000e+00 &0.000e+00& 0.000e+00 &0.000e+00  \\
 6.40 &2.312e-03& 3.555e-03 &0.000e+00& 0.000e+00 &0.000e+00& 0.000e+00 &0.000e+00  \\
 6.45 &2.311e-03& 3.520e-03 &0.000e+00& 0.000e+00 &0.000e+00& 0.000e+00 &0.000e+00   \\
 6.48 &2.312e-03& 3.457e-03 &9.522e-06& 0.000e+00 &0.000e+00& 0.000e+00 &0.000e+00   \\
 6.51 &2.312e-03& 3.341e-03 &1.349e-05& 6.036e-05 &0.000e+00& 0.000e+00 &0.000e+00   \\
 6.54 &2.285e-03& 3.119e-03 &1.678e-05& 1.717e-04 &0.000e+00& 2.484e-05 &0.000e+00   \\
 6.57 &2.225e-03& 2.956e-03 &7.908e-06& 2.263e-04 &0.000e+00& 6.030e-05 &0.000e+00   \\
 6.60 &2.098e-03& 2.814e-03 &1.563e-05& 2.029e-04 &0.000e+00& 1.261e-04 &0.000e+00   \\
 6.63 &1.967e-03& 2.692e-03 &2.299e-05& 1.997e-04 &0.000e+00& 1.323e-04 &0.000e+00   \\
 6.65 &1.897e-03& 2.619e-03 &3.361e-05& 2.296e-04 &0.000e+00& 7.030e-05 &0.000e+00   \\
 6.68 &1.771e-03& 2.484e-03 &7.627e-05& 1.897e-04 &1.852e-06& 1.263e-04 &0.000e+00   \\
 6.70 &1.670e-03& 2.396e-03 &1.524e-04& 8.767e-05 &2.212e-06& 1.021e-04 &0.000e+00   \\
 \hline
\end{tabular}
\label{contributions}
\end{table*}
For each age, given in logarithmic
scale in column 1, we have the number of: massive stars with M$>$20
\Msun in column 2, OB stars in column 3, Wolf-Rayet WN-type stars in
column 4, Wolf-Rayet WC-type stars in column 5, Red Super-Giant (RSG)
in column 6, core collapsed supernovae (SN) in column 7, and Planetary
Nebulae (PN) in column 8.  The general table for the whole grid of IMFs,
ages and metallicities is given in electronic format (see
http://cdsarc.u-strasburg.fr/cats/J.MNRAS.htx).

\begin{figure}
\resizebox{\hsize}{!}{\includegraphics[angle=-90]{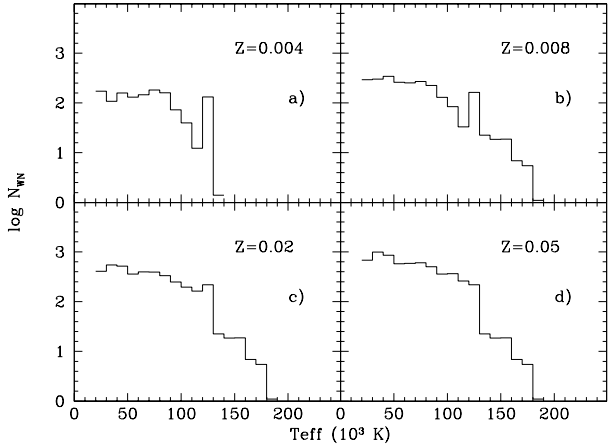}}
\caption{Total number of Wolf-Rayet WN-type stars produced by a
stellar population of 10$^{6}$ M$_{\odot}$ for a STB IMF, as a
function of the hydrostatic temperature of the star. Different panels
correspond to different metallicity, Z: a) 0.004, b) 0.008, c) 0.02
and d) 0.05.}
\label{wn}
\end{figure}
\begin{figure}
\resizebox{\hsize}{!}{\includegraphics[angle=-90]{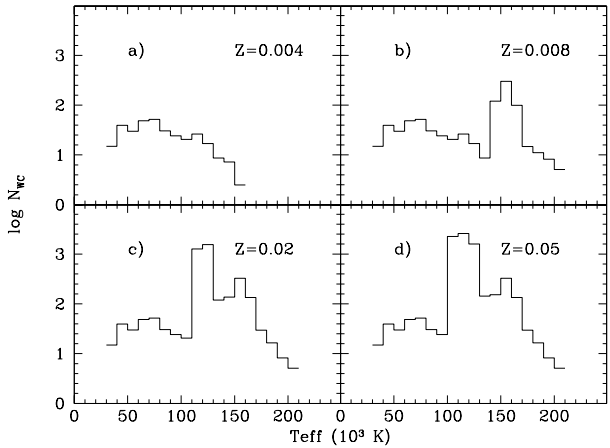}}
\caption{Total number of Wolf-Rayet WC-type stars produced by a
stellar population of 10$^{6}$ M$_{\odot}$ for a STB IMF, as a
function of the hydrostatic temperature of the star. Different panels
correspond to different metallicity, Z: a) 0.004, b) 0.008, c) 0.02
and d) 0.05.}
\label{wc}
\end{figure}

The strong dependence of the star counts in the different phases
with the metallicity is shown in Fig.~\ref{nest_z}.
Here the numbers are normalized to the initial total number
of stars in the isochrone from 0.15~M$_{\odot}$ to 100~M$_{\odot}$.


It can be seen that WR stars appear only for Z $\geq 0.004$. The RSG
number also changes appreciably with metallicity: the number is
smaller for the lowest metallicities but the age range at which they
exist is shorter for higher values of Z. In contrast, OB and PN stars
do not show large variations with the metallicity.

The effects of a different IMF are shown in Fig.~\ref{nest_imf},
limited to the case of $Z = 0.02$.  The slopes of the adopted
IMF affect significantly the number of stars of different types.
For example, the total number of SNs produced by a stellar population
of 1 M$_{\odot}$, is 0.35 10$^{-2}$ for the case FER, 0.53 10$^{-2}$
for KRO, 0.87 10$^{-2}$ for SAL2 and 0.12 10$^{-1}$ for CHA.

The temperature distributions of {\it all} WN and WC stars generated
by an SSP of 10$^6$\Msun\ are shown in Fig.~\ref{wn} and Fig.~\ref{wc},
respectively.  Notice that the temperature is the core-hydrostatic
temperature predicted by stellar evolution tracks and it is not the
one used to assign the atmosphere model, as discussed later.  Notice
also that essentially no WR stars are predicted for Z$ = $0.0001 and
Z$ = $0.0004, in the single stellar evolution scheme.

These histograms have been computed assuming a STB IMF in order to
allow a direct comparison with \citet{snc02}.
The main difference with respect to \citet{snc02}, is that our
predicted number of WN stars, for solar metallicity, is smaller at all
temperatures, but we reach higher temperatures than those ones calculated
with the updated STB99.  The different behavior of the histograms is
due to the different sets of stellar tracks adopted (Padova's for this
work and Geneve's used by STB).

\subsection{Synthesis code description}

To compute the integrated properties of SSPs we have used the
synthesis code by \cite{gmb98}, updated by \cite{mgv00}, together with
a revised version of the atmosphere models and of the nebular emission
contribution.

The main atlas of atmosphere models is from \cite{lcb97}, with its
excellent coverage of effective temperatures, gravities and
metallicities, for stars with Teff$ \leq 25000$ K.  This atlas is used
for all previously defined {\it normal} stars.  For O, B and WR stars
we have taken the NLTE blanketed models by \citet{snc02}, for
Z$ = $0.001, 0.004, 0.008, 0.02 and 0.04. There are 110 models for O-B
stars, with 25000 K $ \rm Teff \leq 51500$ K and $2.95 \leq \log{g}
\leq 4.00$, calculated with the code by \citet*{phl01}, and 120 models
for WR stars (60 WN $+$ 60 WC), calculated with the code CMFGEN by
\cite{hm98}, with 30000 K $\leq T^{*} \leq 120000$ K and
$1.3R_{\odot}\leq R^{*}\leq 20.3 R_{\odot}$ for WN, and with $40000 K
\leq T^{*} \leq 140000$ K and $ 0.8R_{\odot}\leq R^{*}\leq 9.3
R_{\odot}$ for WC. T$^{*}$ and R$^{*}$ are the temperature and the
radius at a Roseland optical depth of 10.

For post-AGB and PN stars we have used the NLTE models by
\cite{rau03}.  In this library the effective temperature goes from
50000 K to 190000 K and $\log{g}$ is between 5.00 and 8.00.  For
higher temperatures we use black bodies.  These models include all
elements from H to Ni and they are available for two values of
metallicities: Z = 0.002 and 0.02. We have used the first
metallicity spectra for our three metal-poor isochrones Z = 0.0001,
0.0004 and 0.004 and the solar abundance spectra for the three other
metallicity isochrones, Z = 0.008, 0.02 and 0.05.

To assign the spectrum to the star along the isochrone we proceed as
follows.  We select the more appropriate
model in Teff and $\log{g}$ in the corresponding spectral library
following the definition of groups of stars.
However, for normal stars with $\rm Teff \ge 63000 $K and an initial
mass $\geq$~15 \Msun\ we use the WN stars models; for normal stars with
initial mass $\leq$ 15~\Msun\ and $\rm Teff \ge 50000 $K or $\rm Teff
\ge 31600$K and gravity $\rm log g > 8$, we use the PN spectral
models, and for WR stars with Teff $<$ 30000~K, we assign a OB spectrum.

\begin{figure}
\resizebox{\hsize}{!}{\includegraphics[angle=-90]{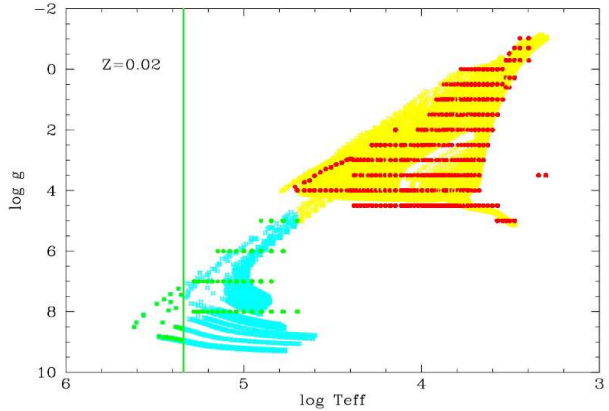}}
\caption{-2. Model assignation for Z = 0.02: yellow and cyan asterisks
correspond to normal stars (NS) and those which end their life as
planetary nebula (PN), respectively according to Padova
isochrones. The available atmosphere models for these stars are shown
over-plotted as red and green solid dots. Notice that the coverage is
good enough. Fig.5-1, 5-3, 5-4, 5-5 and 5-6 corresponding to Z = 0.05,
0.008, 0.004, 0.0004 and 0.0001, respectively, are given in electronic format.}
\label{plano_te_g}
\end{figure}

Fig.~\ref{plano_te_g}-2 shows, for solar abundance, the values of
gravity and effective temperature of the selected stellar models for
normal stars (NS) and for those stars which end their life as
planetary nebula (PN), as red and green dots,
respectively, over-plotted over the effective temperature-gravity
plane of the same kind of stars existing in the H-R diagrams used as
inputs of our code, as yellow and cyan asterisks. The solid green line
at the left part of the figure delimits the upper temperature above
which there is no model for PN stars.  We assign a black body spectrum
to the stars located on the left of this line. The same figures for
the other metallicities are in electronic format.

When the stellar wind is optically thick, as in the case of WR stars,
the isochrone effective temperature cannot be used to assign a stellar
model, since isochrones give the hydrostatic Teff, while atmospheres
use the effective temperature at a Rosseland optical depth of 10,
R$^{*}$. In that case, to assign a model to each WR star, we use the
relationships among optical depth $\tau$, mass loss, $dM/dt$, and wind
velocity, $v(r)$:

\begin{equation}
d\tau=-\kappa(r)\rho(r)dr
\end{equation}
where $\kappa(r) = 0.2(1+X_{\rm S})$ and $X_{\rm S}$, the H surface
abundance by mass (assumed 0.2 for WN and 0 for WC). The mass
conservation requires:
\begin{equation}
\dot{M} = \frac{dM}{dt} = 4\pi r^{2}\rho(r)v(r)dr
\end{equation}
Assuming for the wind velocity the form
$v(r) = v_{\infty}(1-R_{S}/r)^{\beta}$
with  $\beta = 2$ \citep{ber84}, and
integrating these equations, we find the relation between R$^{*}$ and the stellar radius,
R$_{S}$, as:
\begin{equation}
R^{*} =  R_{S}\left[1+\frac{0.02 (1+X_{\rm S})\dot{M}}{4\pi~v_{\infty}~R_{S}} \right]
\end{equation}
This is the radius we use to select the atmosphere model.

The first result of the evolutionary synthesis code is an H-R table
for each age and metallicity, providing the basic quantities along the
isochrones.  As an example, Table ~\ref{HR}, for Z $ = $ \Zsun\ and
age of 0.1 Myr, provides for each stellar mass (column 1), the
effective temperature (column 2), gravity (column 3) and the
luminosity (column 4) of the isochrone, and the values of the
corresponding spectral model in columns 5, 6 and 7,
respectively. Column 8 gives the mass loss rate. The star type is
given in column 9.  The last six columns, 10 to 15, give the number of
stars in each H-R diagram point for each one of our six IMFs.  The
complete set of tables for all ages and metallicities are given in
electronic format.  In the assignation process, the luminosity of the
star along the isochrone is preserved and therefore differences
betweeen column 4 and 7 values are not significant.

\begin{figure}
\resizebox{\hsize}{!}{\includegraphics[angle=0]{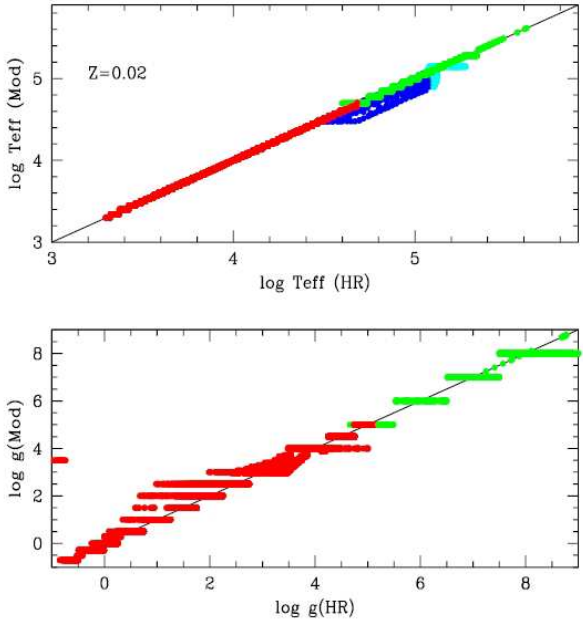}}
\caption{-2. Upper panel: effective temperature assigned to each star as
a function of the corresponding Teff in the isochrone, for solar
abundance as labelled. Bottom panel: same comparison for stellar
gravity values.The other abundances figures are given in electronic
format.}
\label{ajuste}
\end{figure}

Fig.~\ref{ajuste}-2 shows the comparison between the theoretical
(isochrone's) versus the assigned (stellar atmosphere model's) values
for solar abundance. Figures 6-1, 6-3, 6-4, 6-5 and 6-6 corresponding to
Z = 0.05, 0.008, 0.004, 0.0004 and 0.0001 are in electronic format.
The upper panel depicts the assigned effective temperature (stellar
atmosphere model) as a function of the original effective temperature
(isochrone's Teff), in the H-R diagram. The black solid line indicates
the identity. Red dots represent normal stars while green dots are
PNs. In both cases they fall over the identity line.  Wolf Rayet stars
(WC as cyan dots and WN as blue dots, respectively) fall slightly
below this line, which is reasonable since the temperature given in
the stellar models is the temperature at this same Rosseland opacity
equal to 10, lower than the hydrostatic Teff from the isochrone. The
bottom panel shows an equivalent plot but for the gravity (without WR
stars).

\begin{figure}
\resizebox{\hsize}{!}{\includegraphics[angle=0]{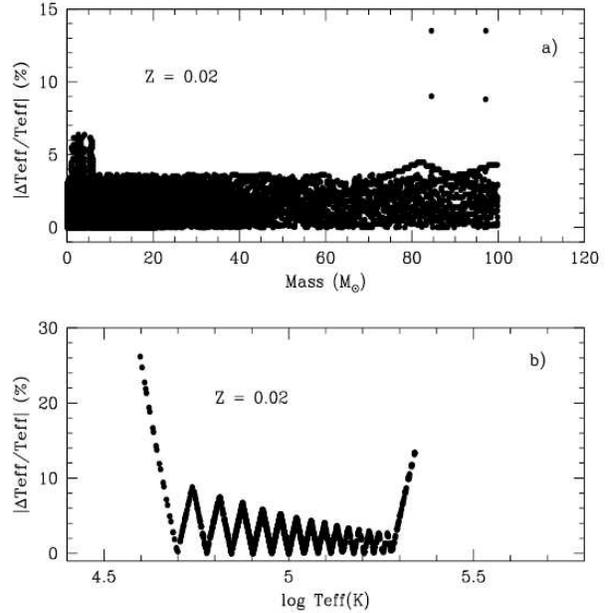}}
\caption{-2. Assignation error, in percentage, in the effective
temperature for solar abundance;
a) normal stars as a function of the their masses and b) post AGB stars (PN), as a function
of Teff.}
\label{ertef}
\end{figure}

Fig.~\ref{ertef}-2 illustrates the assignation errors in the effective
temperature, for solar abundance.  Again the corresponding figures,
7-1 and 7-3 to 7-6 for other metallicities are given in electronic
format.  The error in the assigned effective temperature as a function
of the initial stellar mass, given by the difference between columns
(2) and (5) of the tables, is below 10 per cent for normal stars
(Fig.~\ref{ertef}a), and it is even smaller than 5-7 per cent in most
of them except for some very metal-poor massive stars, for which it
reaches a 15 per cent, and for the lowest mass stars, which are likely
in the border of the PN region. In the first case, the increase in
this error of temperature results from the fact that there are no
atmosphere models for normal stars, hotter than 51000~K. All models
above this temperature correspond to WR stars, but the hottest stars
for $Z = 0.0001 $ and $Z = 0.0004$ are not WR stars and therefore a
normal star atmosphere model with a smaller temperature has been
assigned to them.  For the PN stars (Fig.~\ref{ertef}b) the error, as
a function of the effective temperature, is slightly larger, mostly at
the temperature boundaries.  The main reason is that the temperature
grid for PN atmosphere models is coarser than that of normal stars.

\subsection{Mechanical Energy Injection and associated HII region}

From the number of massive (O, B and WR) stars and SN, we can directly
compute the rate of mechanical energy injected into the interstellar
medium (ISM) by stellar winds and supernova explosions.
At each age, $\tau$, the energy injection rate by stellar winds is:
\begin{equation}
\frac{dE_{W}}{dt}(\tau) = \sum_{M_{*}}{0.5\dot{M}(M_{*})
v_{\inf}^{2}(M_{*})*n_{WR}(M_{*})}
\label{eq:wind}
\end{equation}
where M$_{*}$, $n_{WR}(M_{*})$ and $\dot{M}$ are the initial mass and
number along the isochrone, and the current mass loss rate,
respectively.
The energy injection rate is obtained after
adding to Eq. \ref{eq:wind} the energy contribution by the
SN  rate
\begin{equation}
\frac{dE}{dt}(\tau) = \frac{dE_{W}}{dt} + \epsilon_{SN}\frac{dn_{SN}}{dt}
\label{eq:erate}
\end{equation}
where $\epsilon_{SN}$ is the  energy injected by a single SN event and
$\frac{dn_{SN}}{dt} = \nu_{SN}(\tau)$ is the SN rate or number of SN in each
time step, given in Table~\ref{contributions}.


We can derive the radius of the associated H{\sc ii} region from the
mechanical energy produced by the expanding atmospheres of massive
stars with strong winds.  \cite{cmw75} demonstrated that an early-type
star with a strong stellar wind can blow out a large cavity or
bubble in the surrounding gas, if it is assumed to be compressed
into a thin spherical shell. The wind-driven shell begins to evolve
with an initial phase of free expansion followed
\onecolumn
\landscape
\begin{table*}
\caption{H-R diagram for stars with 50\Msun\ $\rm \leq M \leq 75$
  \Msun\ and \Zsun\ and an age of 0.1 Myr ($\log{Age}$ = 5.00). The
  whole table is available in electronic format together with the
  complete set of H-R diagrams.}
\begin{tabular}{ccccccccccccccc}
\hline
Mass$_{*}$ HR &  Teff-HR & $\log{g}$-HR & $\log{L}$-HR & Teff-Mod &  $\log{g}$-Mod & $\log{L}$-Mod & log(Mdot)&  WR & N$_{*}$ SAL1 & N$_{*}$ SAL2 & N$_{*}$ FER &  N$_{*}$ KRO &  N$_{*}$ CHA &  N$_{*}$ STB \\
 \Msun & K  &   & \Lsun &  K  &   & \Lsun &  & & & & & & \\
\hline
 50.30 &44771.33 & 4.13 &   5.57& 50000. & 4.00. & 5.60 & -5.903&  NS& 2.020e-05 &1.010e-05& 1.960e-06 &4.050e-06& 1.520e-05& 2.190e-05   \\
 50.80 &44874.54 & 4.13 &   5.58& 50000. & 4.00. & 5.61 & -5.892&  NS& 1.970e-05 &9.840e-06& 1.900e-06 &3.940e-06& 1.480e-05& 2.140e-05   \\
 51.30 &44977.97 & 4.13 &   5.59& 50000. & 4.00. & 5.61 & -5.881&  NS& 1.920e-05 &9.620e-06& 1.850e-06 &3.840e-06& 1.450e-05& 2.090e-05   \\
 51.80 &45081.65 & 4.12 &   5.60& 50000. & 4.00. & 5.62 & -5.870&  NS& 1.880e-05 &9.400e-06& 1.790e-06 &3.740e-06& 1.420e-05& 2.050e-05   \\
 52.30 &45185.62 & 4.12 &   5.61& 50000. & 4.00. & 5.62 & -5.860&  NS& 1.840e-05 &9.190e-06& 1.740e-06 &3.640e-06& 1.390e-05& 2.000e-05   \\
 52.80 &45185.62 & 4.12 &   5.61& 50000. & 4.00. & 5.63 & -5.849&  NS& 1.800e-05 &8.990e-06& 1.700e-06 &3.550e-06& 1.360e-05& 1.960e-05   \\
 53.30 &45289.77 & 4.12 &   5.62& 50000. & 4.00. & 5.64 & -5.839&  NS& 1.760e-05 &8.790e-06& 1.650e-06 &3.460e-06& 1.330e-05& 1.910e-05   \\
 53.80 &45394.17 & 4.12 &   5.63& 50000. & 4.00. & 5.64 & -5.828&  NS& 1.720e-05 &8.600e-06& 1.600e-06 &3.380e-06& 1.300e-05& 1.870e-05   \\
 54.30 &45498.80 & 4.12 &   5.64& 50000. & 4.00. & 5.65 & -5.818&  NS& 1.680e-05 &8.410e-06& 1.560e-06 &3.290e-06& 1.270e-05& 1.830e-05   \\
 54.80 &45603.68 & 4.12 &   5.65& 50000. & 4.00. & 5.65 & -5.808&  NS& 1.650e-05 &8.230e-06& 1.520e-06 &3.210e-06& 1.240e-05& 1.790e-05   \\
 55.30 &45708.80 & 4.12 &   5.66& 50000. & 4.00. & 5.66 & -5.798&  NS& 1.610e-05 &8.060e-06& 1.480e-06 &3.130e-06& 1.220e-05& 1.760e-05   \\
 55.80 &45708.80 & 4.11 &   5.67& 50000. & 4.00. & 5.66 & -5.788&  NS& 1.580e-05 &7.890e-06& 1.440e-06 &3.060e-06& 1.190e-05& 1.720e-05   \\
 56.30 &45814.16 & 4.11 &   5.68& 50000. & 4.00. & 5.67 & -5.778&  NS& 1.550e-05 &7.730e-06& 1.400e-06 &2.990e-06& 1.170e-05& 1.680e-05   \\
 56.80 &45919.82 & 4.11 &   5.68& 50000. & 4.00. & 5.67 & -5.768&  NS& 1.510e-05 &7.570e-06& 1.370e-06 &2.920e-06& 1.150e-05& 1.650e-05   \\
 57.30 &46025.67 & 4.11 &   5.69& 50000. & 4.00. & 5.68 & -5.758&  NS& 1.480e-05 &7.420e-06& 1.330e-06 &2.850e-06& 1.120e-05& 1.620e-05   \\
 57.80 &46131.76 & 4.11 &   5.70& 50000. & 4.00. & 5.68 & -5.748&  NS& 1.450e-05 &7.260e-06& 1.300e-06 &2.780e-06& 1.100e-05& 1.580e-05   \\
 58.30 &46238.10 & 4.11 &   5.71& 50000. & 4.00. & 5.69 & -5.738&  NS& 1.420e-05 &7.120e-06& 1.270e-06 &2.720e-06& 1.080e-05& 1.550e-05   \\
 58.80 &46238.10 & 4.10 &   5.72& 50000. & 4.00. & 5.70 & -5.729&  NS& 1.400e-05 &6.980e-06& 1.240e-06 &2.660e-06& 1.060e-05& 1.520e-05   \\
 59.30 &46344.68 & 4.10 &   5.73& 50000. & 4.00. & 5.70 & -5.719&  NS& 1.370e-05 &6.840e-06& 1.210e-06 &2.600e-06& 1.040e-05& 1.490e-05   \\
 59.80 &46451.51 & 4.10 &   5.74& 50000. & 4.00. & 5.71 & -5.710&  NS& 1.340e-05 &6.710e-06& 1.180e-06 &2.540e-06& 1.020e-05& 1.460e-05   \\
 60.30 &46558.63 & 4.10 &   5.74& 50000. & 4.00. & 5.71 & -5.702&  NS& 1.320e-05 &6.580e-06& 1.150e-06 &2.480e-06& 9.990e-06& 1.430e-05   \\
 60.80 &46558.63 & 4.10 &   5.75& 50000. & 4.00. & 5.71 & -5.697&  NS& 1.290e-05 &6.450e-06& 1.120e-06 &2.430e-06& 9.800e-06& 1.400e-05   \\
 61.30 &46558.63 & 4.10 &   5.75& 50000. & 4.00. & 5.72 & -5.691&  NS& 1.270e-05 &6.330e-06& 1.090e-06 &2.370e-06& 9.620e-06& 1.380e-05   \\
 61.80 &46558.63 & 4.10 &   5.76& 50000. & 4.00. & 5.72 & -5.686&  NS& 1.240e-05 &6.210e-06& 1.070e-06 &2.320e-06& 9.440e-06& 1.350e-05   \\
 62.30 &46665.95 & 4.10 &   5.76& 50000. & 4.00. & 5.72 & -5.680&  NS& 1.220e-05 &6.090e-06& 1.040e-06 &2.270e-06& 9.270e-06& 1.330e-05   \\
 62.80 &46665.95 & 4.10 &   5.77& 50000. & 4.00. & 5.73 & -5.675&  NS& 1.200e-05 &5.980e-06& 1.020e-06 &2.220e-06& 9.100e-06& 1.300e-05   \\
 63.30 &46665.95 & 4.10 &   5.77& 50000. & 4.00. & 5.73 & -5.669&  NS& 1.170e-05 &5.870e-06& 9.950e-07 &2.180e-06& 8.940e-06& 1.280e-05   \\
 63.80 &46665.95 & 4.10 &   5.78& 50000. & 4.00. & 5.74 & -5.664&  NS& 1.150e-05 &5.760e-06& 9.730e-07 &2.130e-06& 8.780e-06& 1.250e-05   \\
 64.30 &46773.52 & 4.10 &   5.78& 50000. & 4.00. & 5.74 & -5.658&  NS& 1.130e-05 &5.660e-06& 9.500e-07 &2.090e-06& 8.620e-06& 1.230e-05   \\
 64.80 &46773.52 & 4.10 &   5.79& 50000. & 4.00. & 5.74 & -5.653&  NS& 1.110e-05 &5.550e-06& 9.290e-07 &2.040e-06& 8.470e-06& 1.210e-05   \\
 65.30 &46773.52 & 4.10 &   5.79& 50000. & 4.00. & 5.75 & -5.647&  NS& 1.090e-05 &5.450e-06& 9.080e-07 &2.000e-06& 8.320e-06& 1.190e-05   \\
 65.80 &46881.34 & 4.10 &   5.80& 50000. & 4.00. & 5.75 & -5.641&  NS& 1.070e-05 &5.360e-06& 8.880e-07 &1.960e-06& 8.180e-06& 1.170e-05   \\
 66.30 &46881.34 & 4.10 &   5.80& 50000. & 4.00. & 5.76 & -5.636&  NS& 1.050e-05 &5.260e-06& 8.690e-07 &1.920e-06& 8.040e-06& 1.150e-05   \\
 66.80 &46881.34 & 4.10 &   5.81& 50000. & 4.00. & 5.76 & -5.631&  NS& 1.030e-05 &5.170e-06& 8.500e-07 &1.880e-06& 7.900e-06& 1.130e-05   \\
 67.30 &46989.40 & 4.09 &   5.81& 50000. & 4.00. & 5.76 & -5.625&  NS& 1.020e-05 &5.080e-06& 8.310e-07 &1.840e-06& 7.760e-06& 1.110e-05   \\
 67.80 &46989.40 & 4.09 &   5.82& 50000. & 4.00. & 5.77 & -5.619&  NS& 9.990e-06 &4.990e-06& 8.130e-07 &1.810e-06& 7.630e-06& 1.090e-05   \\
 68.30 &46989.40 & 4.09 &   5.82& 50000. & 4.00. & 5.77 & -5.614&  NS& 9.820e-06 &4.910e-06& 7.960e-07 &1.770e-06& 7.500e-06& 1.070e-05   \\
 68.80 &46989.40 & 4.09 &   5.83& 50000. & 4.00. & 5.78 & -5.609&  NS& 9.650e-06 &4.820e-06& 7.790e-07 &1.740e-06& 7.380e-06& 1.050e-05   \\
 69.30 &47097.72 & 4.09 &   5.83& 50000. & 4.00. & 5.78 & -5.603&  NS& 9.490e-06 &4.740e-06& 7.630e-07 &1.700e-06& 7.260e-06& 1.030e-05   \\
 69.80 &47097.72 & 4.09 &   5.84& 50000. & 4.00. & 5.78 & -5.597&  NS& 9.330e-06 &4.660e-06& 7.470e-07 &1.670e-06& 7.140e-06& 1.020e-05   \\
 70.30 &47097.72 & 4.09 &   5.84& 50000. & 4.00. & 5.79 & -5.592&  NS& 9.170e-06 &4.590e-06& 7.310e-07 &1.640e-06& 7.020e-06& 9.980e-06   \\
 70.80 &47206.28 & 4.09 &   5.85& 50000. & 4.00. & 5.79 & -5.586&  NS& 9.020e-06 &4.510e-06& 7.160e-07 &1.610e-06& 6.910e-06& 9.820e-06   \\
 71.30 &47206.28 & 4.09 &   5.85& 50000. & 4.00. & 5.79 & -5.581&  NS& 8.870e-06 &4.440e-06& 7.010e-07 &1.580e-06& 6.800e-06& 9.660e-06   \\
 71.80 &47206.28 & 4.09 &   5.86& 50000. & 4.00. & 5.80 & -5.575&  NS& 8.730e-06 &4.360e-06& 6.870e-07 &1.550e-06& 6.690e-06& 9.500e-06   \\
 72.30 &47206.28 & 4.09 &   5.86& 50000. & 4.00. & 5.80 & -5.570&  NS& 8.590e-06 &4.290e-06& 6.730e-07 &1.520e-06& 6.580e-06& 9.350e-06   \\
 72.80 &47315.15 & 4.09 &   5.86& 50000. & 4.00. & 5.80 & -5.564&  NS& 8.450e-06 &4.220e-06& 6.600e-07 &1.490e-06& 6.480e-06& 9.200e-06   \\
 73.30 &47315.15 & 4.09 &   5.87& 50000. & 4.00. & 5.81 & -5.559&  NS& 8.320e-06 &4.160e-06& 6.470e-07 &1.460e-06& 6.380e-06& 9.050e-06   \\
 73.80 &47315.15 & 4.09 &   5.87& 50000. & 4.00. & 5.81 & -5.553&  NS& 8.180e-06 &4.090e-06& 6.340e-07 &1.440e-06& 6.280e-06& 8.910e-06   \\
 74.30 &47424.21 & 4.09 &   5.88& 50000. & 4.00. & 5.81 & -5.548&  NS& 8.050e-06 &4.030e-06& 6.210e-07 &1.410e-06& 6.180e-06& 8.770e-06   \\
 74.80 &47424.21 & 4.09 &   5.88& 50000. & 4.00. & 5.82 & -5.542&  NS& 7.930e-06 &3.960e-06& 6.090e-07 &1.390e-06& 6.090e-06& 8.630e-06   \\
\hline
\end{tabular}
\label{HR}
\end{table*}
\endlandscape
\twocolumn
 by an adiabatic
expansion phase, and then the material collapses into a thin, cold
shell as a result of radiative cooling. 
At this stage the gas traps
the ionization front and the radiative phase begins. In this phase the
ionizing photons are absorbed and the region cools via emission in the
Balmer lines. In this process, the radius of the outer shock, $\rm
R_{HII Reg}$ also called R$_{s}$, evolves as:

\begin{eqnarray}
R_{HII Reg} = 1.6(\epsilon/n)^{1/5} \tau^{3/5}  (pc)
\end{eqnarray}

where $\epsilon=\frac{dE}{dt}(\tau) $ is the total mechanical energy (SN
and stellar winds) per unit time injected in units of 10$^{36}$ ergs
s$^{-1}$, calculated with Eq. (9), \textit{n} is the interstellar
medium density in units of cm$^{-3}$, and \textit{$\tau$} the age of the
shell in units of 10$^{4}$ yr.
We have extrapolated this bubble geometry to a shell
structure formed by the combined effects of the mechanical energy
deposition from the winds coming from massive stars in the ionizing
cluster and from SN explosions. Then, the ionized gas is assumed to be
located in a thin spherical shell at that distance from the ionizing
source.

Table~\ref{emec} gives the results of these calculations.  We show for
each age (column 1) the mechanical energy corresponding to the
supernova explosions (column 2) same from stellar winds (column 2),
and total (column 4). The radius, in pc, of the H{\sc ii} region
produced by this energy and calculated as described above, is given in
column 5. The energy scales with the mass of the stellar cluster and
therefore the radius of the HII region depends on this mass as
$M^{1/5}$, which must be taken into account when using this table.  We
show here only the results for the first 5 Myr of the evolution of a
stellar cluster of 1 \Msun\ with IMF SAL2 and solar metallicity.  The
complete set of tables for all ages, metallicities and IMFs are given
in electronic format.
\begin{table}
\caption{Mechanical energy injected into the ISM by a stellar
cluster of 1 M$_{\odot}$ due to massive star winds and to supernova
explosions, and the corresponding H{\sc ii} region radius for the
first 5 Myr, an IMF SAL2, and solar metallicity. The whole table is
available in electronic format.}
\begin{tabular}{ccccc}
\hline
 Log(Age) &  Emec$_{\rm SN}$ &  Emec$_{\rm Wind}$ &   Emec$_{tot}$ & R$_{H{\sc ii}reg}$ \\
 (yr)   & \multicolumn{3}{c}{($10^{47}$ erg)} & (pc)\\
 5.00 &  0.0000 &  0.0968 &  0.0968  &   1.26 \\
 5.48 &  0.0000 &  0.2613 &  0.2613  &   2.39 \\
 5.70 &  0.0000 &  0.4193 &  0.4193  &   3.22 \\
 5.85 &  0.0000 &  0.6800 &  0.6800  &   4.06 \\
 6.00 &  0.0000 &  1.2113 &  1.2113  &   5.26 \\
 6.11 &  0.0000 &  2.2210 &  2.2210  &   6.59 \\
 6.18 &  0.0000 &  2.4558 &  2.4558  &   7.12 \\
 6.23 &  0.0000 &  2.5839 &  2.5839  &   7.57 \\
 6.30 &  0.0000 &  3.3025 &  3.3025  &   8.48 \\
 6.34 &  0.0000 &  3.1173 &  3.1173  &   8.71 \\
 6.40 &  0.0000 &  3.8874 &  3.8874  &   9.58 \\
 6.45 &  0.0000 &  4.6465 &  4.6465  &  10.39 \\
 6.48 &  0.0000 &  4.4463 &  4.4463  &  10.59 \\
 6.51 &  0.0000 &  8.9305 &  8.9305  &  12.49 \\
 6.54 &  0.2484 & 24.3092 & 24.5576  &  15.85 \\
 6.57 &  0.6030 & 34.6753 & 35.2783  &  17.42 \\
 6.60 &  1.2607 & 38.7365 & 39.9972  &  18.43 \\
 6.63 &  1.3229 & 41.0796 & 42.4025  &  19.19 \\
 6.65 &  0.7030 & 51.0340 & 51.7370  &  20.34 \\
 6.68 &  1.2626 & 52.8720 & 54.1346  &  21.06 \\
 6.70 &  1.0206 & 33.6511 & 34.6717  &  19.58 \\
\hline
\end{tabular}
\label{emec}
\end{table}

\subsection{Stellar Spectral Energy Distributions}

We have obtained the Spectral Energy Distributions (SED's) of simple
stellar populations for different ages (from $\log{\tau}  = 5.00$ to
10.18), metallicities (Z$ = $0.0001, 0.004, 0.004, 0.008, 0.02, 0.05)
and for several assumptions concerning the IMF (Table~\ref{imfs}).
\begin{figure}
\resizebox{\hsize}{!}{\includegraphics[angle=-90]{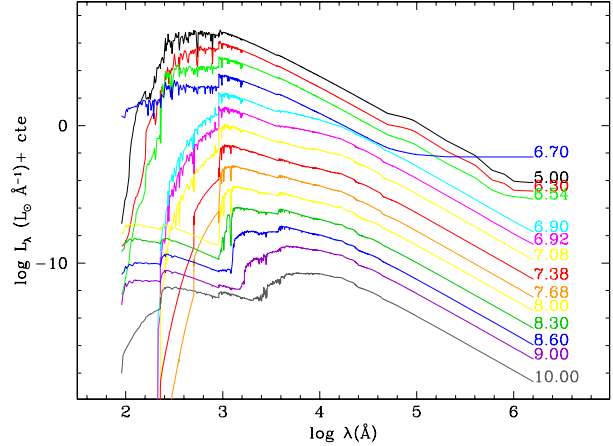}}
\caption{Stellar Spectral Energy Distributions for the SAL2 IMF and
for \Zsun\ at the ages labelled in the panel}
\label{sps}
\end{figure}
\begin{figure}
\resizebox{\hsize}{!}{\includegraphics[angle=0]{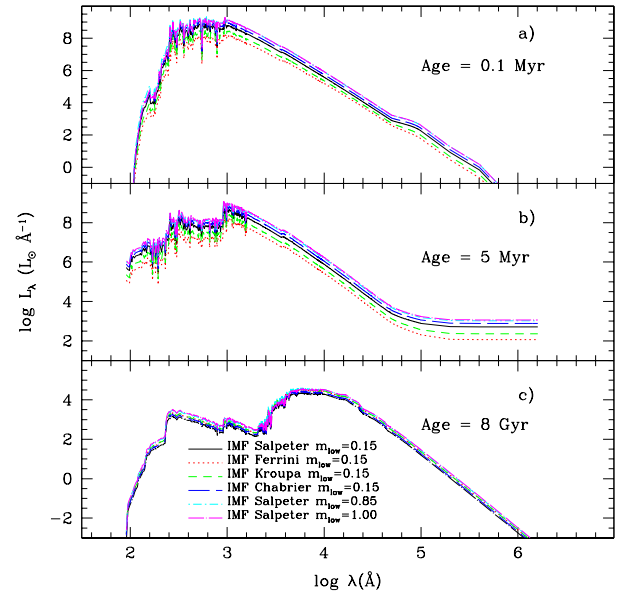}}
\caption{Comparison of the resulting SED obtained with the 6
IMFs  for an age of a) 0.1 Myr; b) 5 Myr and c) 8 Gyr}
\label{sp-imf}
\end{figure}
Fig.~\ref{sps} shows spectra for selected ages for Z $ = $ \Zsun\ and
adopting the SAL2 IMF. The new SEDs are less hard than our previous
models \citep{gmb98,mgv00} because of the use of the NLTE model
atmospheres of massive stars.
Above 54ev the emergent flux is determined by the wind density as a
function of metallicity and, in the NLTE models, the hardness of
ionizing radiation decreases. The models produce a lower flux in the
HeI continuum, for Z $>$ 0.4 \Zsun\ and Age $<$ 7 Myr. The consequence
of using the new models will be a lower excitation, in HII regions
ionized by young clusters.  Our previous models, with hotter
atmospheres for massive stars \citep{gbd95a}, could explain the
emission line ratios observed in high metallicity H{\sc ii} regions,
only assuming a steepening of the IMF (i.e. a smaller number of
massive stars) or invoking mass segregation effects in small
clusters. Today, we know that this is against evidence since HST
discovered very massive stars also in small clusters. The new models
will explain these observations in a more natural way, as we will show
in Paper II.

\begin{figure*}
\resizebox{\hsize}{!}{\includegraphics[angle=-90]{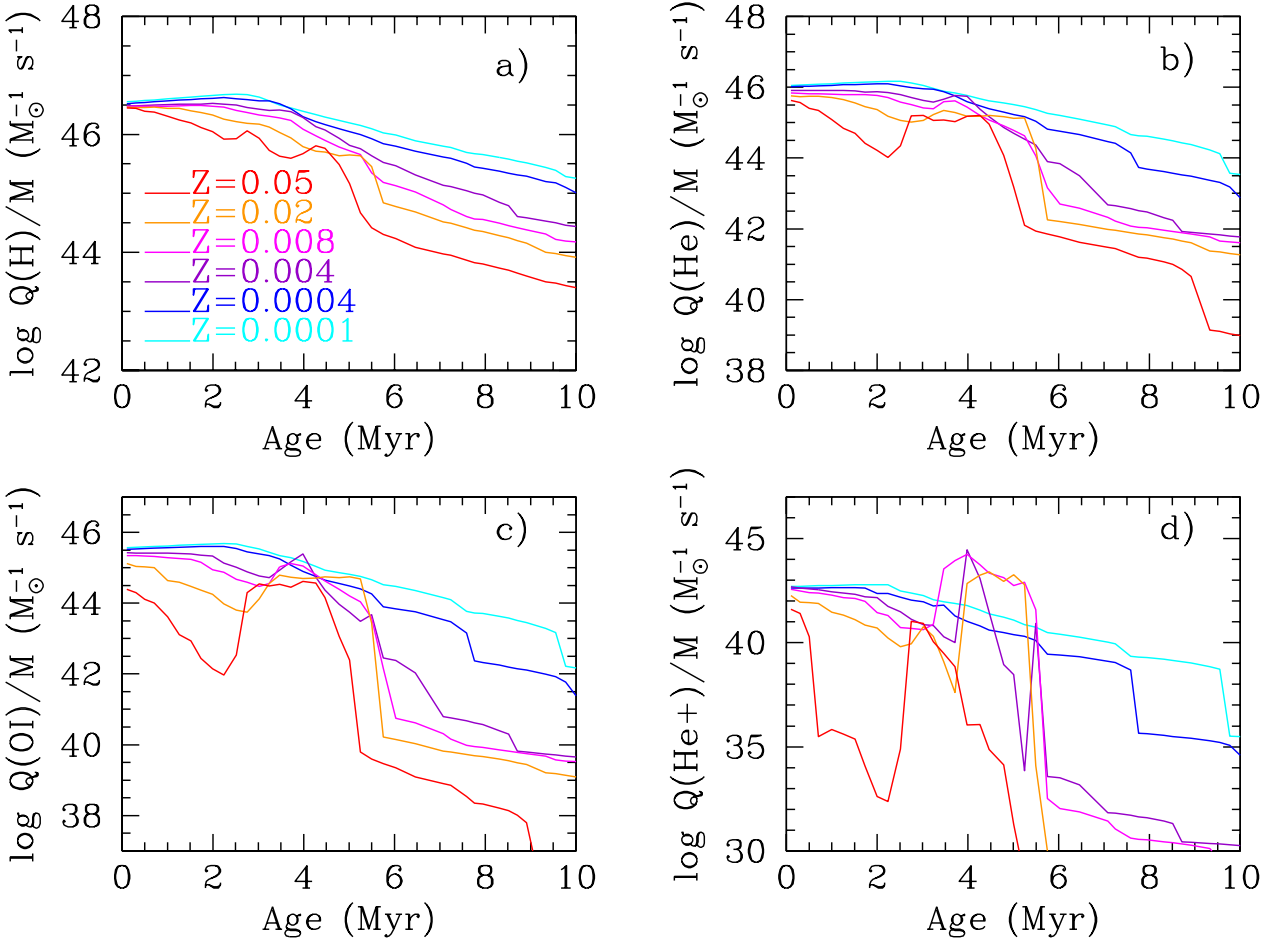}}
\caption{Time evolution of the number of ionizing photons per cluster
solar mass, for SAL2 and six metallicities, for the first 5 Myr
evolution. Different panels show different ionization photons
evolution, labelled as: a) Q(H), b) Q(He), c) Q(OI) and d)
Q(He$^{+}$).}
\label{photons}
\end{figure*}
On the other hand, when Planetary Nebulae appear between $\log{\tau} =
8.30$ and $\log{\tau} = 10.00 $, the SEDs are harder than in our
previous models. This is due to the new models from \cite{rau03},
especially computed for these very hot stellar evolution phases.

Differences due to the adopted IMF are shown in Fig.~\ref{sp-imf} for
three selected ages and for \Zsun\. They may be significant because
of the different contribution of massive stars in the young isochrones
and of low mass stars in the old isochrones.

\section{Nebular plus stellar spectral energy distributions}

In this section we describe how we include the the H and He nebular
continua, free-free, free-bound and two-photons emission mechanisms,
to the SED of simple stellar populations.

\subsection{Nebular contribution calculation}

We first compute the total number of ionizing photons, Q(H), Q(HeI),
Q(HeII) and Q(OI), by integrating the specific photon luminosity of a
given SED from $\lambda = 0$ to 912\AA, 504\AA, 228\AA\ and 353.3\AA\,
respectively. The evolution of the numbers of ionizing photons for the
6 metallicities and the 6 IMFs is given in electronic
format. Table~\ref{qh} shows as an example the results for the SAL2
IMF and Z = Z$_{\odot}$ and for the first 5 Myr. For each age, (column
1), we provide the the number of ionizing photons, Q(HeII), Q(OI),
Q(HeI) and Q(H), in columns 2 to 5, respectively.  These quantities
are also plotted in Fig.~\ref{photons} for the first 5 Myr.  These
numbers are higher at low than at high metallicities, and show a smooth
evolution with age.  At the highest metallicities there are abrupt
variations, especially for Q(HeII) and Q(OI), syncronized with the
appearance of the WR phase.

\begin{figure}
\resizebox{\hsize}{!}{\includegraphics[angle=-90]{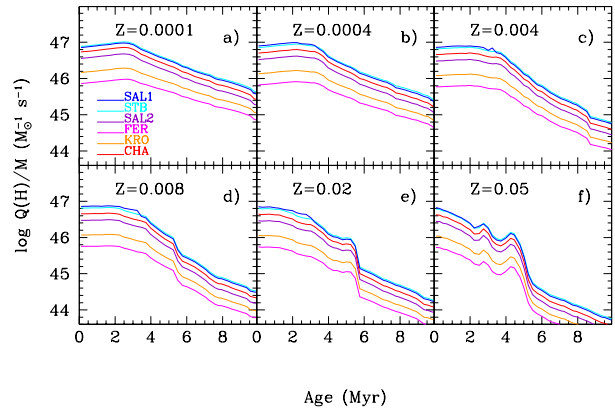}}
\caption{Time evolution for the first 5 Myr of the number of ionizing
photons Q(H) for the six IMFs used in this work: a) Z $ = $ 0.0001,
b) Z $ = $ 0.0004, c) Z $ = $ 0.004, d) Z $ = $ 0.008, e) Z $ = $ 0.02, and f)
Z $ = $ 0.05.}
\label{qh_imf}
\end{figure}

Effects of different IMF are shown in Fig.~\ref{qh_imf} limited to the
hydrogen ionizing photon flux.  The differences mirror the different
number of massive stars for a given cluster initial mass, and its
different evolution with time.

Once Q(H) has been computed, we can calculate the nebular
contribution. We have included the hydrogen and helium (both He and
He$^{+}$) free-free and free-bound emission processes as well as the
2-photon continuum.  We have used the expression from \cite{ost89}:
\begin{equation}
L_{\lambda} (ers.s^{-1}.\AA^{-1}.M_{\odot}^{-1}) = \Gamma
\times\frac{c}{\lambda^{2}.\alpha_{B}(HI)}Q(H)
\end{equation}
where $Q(H)$ is the number of ionizing photons, $c$ is the light
velocity, and $\alpha_{B}$ is the recombination coefficient to the excited
level in hydrogen, which depends on the electronic temperature.  For the latter
we have used values that depend
on the metallicity, as  summarized in Table~\ref{tem}.

\begin{table}
\caption{The number of ionizing photons Q(HeII), Q(OI),
 Q(HeI) and Q(H) emitted by a 1 \Msun\ stellar cluster with an
 IMF SAL2, and solar metallicity during the first 5 Myr. The whole
 table is available in electronic format.}
\begin{tabular}{ccccc}
\hline
$\log{Age}$ & $\log{Q(HeII)}$ & $\log{Q(OI)} $  & $\log{Q(HeI)}$ & $\log{Q(H)}$\\
  (yr)      &    $s^{-1}$     &  $s^{-1}$   &    $s^{-1}$   &  $s^{-1}$ \\
\hline
 5.00 & 42.26 & 45.12 & 45.76 & 46.46  \\
 5.48 & 41.94 & 45.03 & 45.74 & 46.45  \\
 5.70 & 41.92 & 45.02 & 45.74 & 46.46  \\
 5.85 & 41.89 & 45.00 & 45.74 & 46.46  \\
 6.00 & 41.46 & 44.64 & 45.70 & 46.44  \\
 6.10 & 41.32 & 44.59 & 45.65 & 46.44  \\
 6.18 & 41.11 & 44.48 & 45.56 & 46.40  \\
 6.24 & 40.85 & 44.37 & 45.46 & 46.37  \\
 6.30 & 40.70 & 44.25 & 45.37 & 46.33  \\
 6.35 & 40.23 & 43.98 & 45.18 & 46.26  \\
 6.40 & 39.81 & 43.80 & 45.07 & 46.21  \\
 6.44 & 39.94 & 43.75 & 45.01 & 46.19  \\
 6.48 & 40.77 & 44.12 & 45.07 & 46.18  \\
 6.51 & 40.30 & 44.55 & 45.23 & 46.12  \\
 6.54 & 39.09 & 44.80 & 45.34 & 46.03  \\
 6.57 & 37.61 & 44.73 & 45.28 & 45.94  \\
 6.60 & 42.84 & 44.70 & 45.19 & 45.80  \\
 6.63 & 43.24 & 44.71 & 45.18 & 45.72  \\
 6.65 & 43.40 & 44.74 & 45.21 & 45.70  \\
 6.68 & 42.94 & 44.73 & 45.17 & 45.64  \\
 6.70 & 43.26 & 44.74 & 45.12 & 45.65  \\
 \hline
\end{tabular}
\label{qh}
\end{table}

The function $\Gamma$ is the sum of the emission coefficients for Hydrogen
and Helium, including both free-free and free-bound contributions, and
the emission coefficient due to the two-photons continuum. That is:
$\Gamma = \Gamma(HI)+\Gamma(HeI)\times\frac{N(HeII)}{N(H)}+\Gamma(2q)$
The last function $\Gamma(2q)$ coefficient is also taken from \cite{ost89}:
\begin{equation}
\Gamma(2q) = \frac{\alpha_{eff}g_{\nu}}{1+q_{2}/A_{2q}}
\end{equation}
with the coefficient $A_{2q} = 8.2249$. The values of $\alpha_{eff}$ and
$q_{2}$ also depend on the electronic temperature
(Table~\ref{tem}). We have used the values given by \cite{ost89} and
we have performed a linear fitting, obtaining:
\begin{equation}
\alpha_{eff} = 0.647 10^{-10} Te^{-0.722}
\end{equation}
and
\begin{equation}
q_{2} = 5.92.10^{-4}-6.1.10^{-9}Te
\end{equation}

The function $g_{\nu}$ is taken from \cite{nuss84}.

\begin{table}
\caption{Electronic temperature assigned to each metallicity Z}
\begin{tabular}{lr}
Z & Te \\
  & (K) \\
\hline
0.0001 & 19950 \\
0 0004 & 15850 \\
0.004  & 10000  \\
0.008  & 7940 \\
0.02   & 6310 \\
0.05   & 3160 \\
\hline
\end{tabular}
\label{tem}
\end{table}
The functions $\Gamma(HI)$ and $\Gamma(HeI)$ have two terms, due to
free-free and free-bound emissions. The terms due to free-free
emission are calculated with the classical expressions from
\cite{bro70} and \cite{ost89}.

\begin{figure}
\resizebox{\hsize}{!}{\includegraphics[angle=-90]{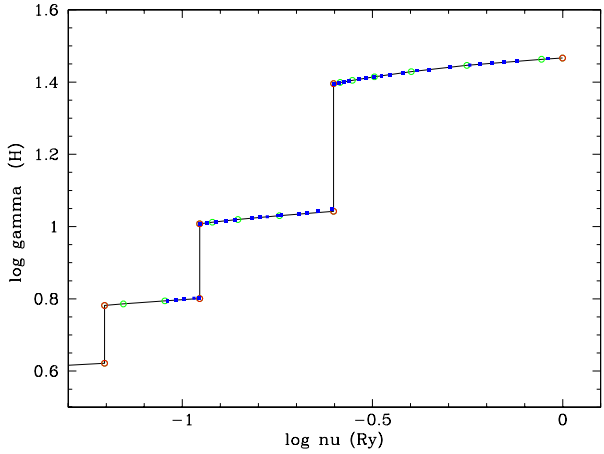}}
\caption{$\Gamma^{+}$ {\sl vs} function the frequency $\nu$ (in Ryds),
for the hydrogen free-bound nebular emission given by \citet{erc06}
as green and red open dots, and the points obtained with our interpolation,
blue full squares, to the wavelengths used in our SEDs.}
\label{checking_h}
\end{figure}

The free-bound contributions have been computed using the recent work
by \cite{erc06}. These last authors give the coefficients for the
calculations of the continuous emission spectra of HI, HeI and HeII
for energies from the first ionization threshold to the $n = 20$
threshold of hydrogen and for electronic temperatures in the range
$100 \le Te \le 10^{5}$ K.  Their results are given in tables where
they tabulate the function:
\begin{equation}
\Gamma_{ff}^{+}(\nu) = \Gamma_{ff}(\nu)10^{34}T^{3/2}
e^{\Delta E/kT} = \Gamma_{ff}(\nu)10^{40}t^{3/2}e^{15.7887\Delta E_{R}/t}
\label{eq:gamma}
\end{equation}
being $t = T(K)/10^{4}$, $\Delta E$ is the difference between the photon
energy and the energy of the nearest threshold of the lower energy and
$\Delta E_{R}$ is the same energy difference expressed in Ryd units.
We have  computed the values of the function $\Gamma_{ff}^{+}(\nu)$
for HI, HeI and HeII by interpolating in wavelength, and
for the electronic temperatures corresponding to the adopted metallicities
given in Table~\ref{tem}. Fig.~\ref{checking_h} shows the $\Gamma^{+}$
function, used for the hydrogen. Green and red empty circles are the
values tabulated by the above authors, while the filled blue squares are
the values obtained for our wavelengths. Then, we use Eq.\ref{eq:gamma} to
compute $\Gamma_{ff}(\nu)$.

\begin{figure}
\resizebox{\hsize}{!}{\includegraphics[angle=-90]{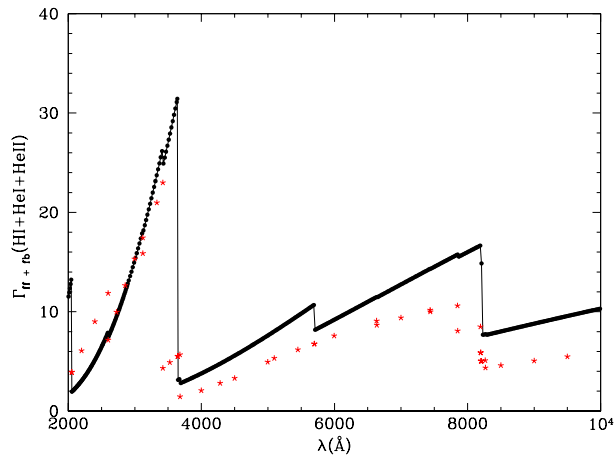}}
\caption{Free-free and free-bound nebular emission gamma function for
H and He obtained with the updated calculations for an electronic
temperature of 10000 K (solid black dots), compared with the old
values, plotted as red stars.}
\label{gamma_chequeo}
\end{figure}

\begin{figure}
\resizebox{\hsize}{!}{\includegraphics[angle=0]{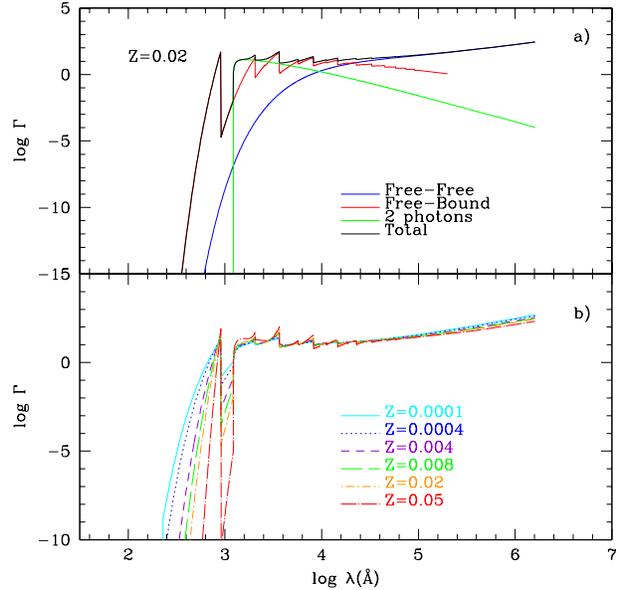}}
\caption{The total $\Gamma$ function solid line, as the sum of the
three contributions of nebular continuum. Panel a) displays the three
contributions: 1) free-free emission due to H and He, blue dots,
2) free-bound emission, red squares, and 3) two-photons contribution, green
stars, for Z$ = 0.02$. Panel b) displays the total function for the six
metallicities computed in this work as labelled in the plot.}
\label{gamma}
\end{figure}

From these coefficients and those for the free-free emission we have
obtained the total $\Gamma_{ff+fb}$(HI+HeI+HeII). This total function
is slightly different than the one we used in \cite{mgv00} taken from
\cite{aller}.  Fig.~\ref{gamma_chequeo} compares the new results (black dots)
with the old results (red stars), for $T = 10000$~K.
The total function $\Gamma$ is then found when we add the two-photons
contribution.  This total function is shown in Fig.~\ref{gamma} for
Z = Z$_\odot$. Panel a) displays the three nebular contributions, free-free,
free-bound and two-photons, plotted as filled blue  dots, red open
squares and green stars, respectively. This function depends on the
electronic temperature, and therefore changes with metallicity. Panel
b) displays the function for the six metallicities used in this work.
The metallicity affects this function only below
$\lambda \sim2000 \AA$.

\subsection{Luminosity and Equivalent Widths for \halpha\ and \hbeta }

We have computed the intensity of the Balmer emission lines, in
particular I$_{H_{\alpha}}$ and I$_{H_{\beta}}$, with the
following equations from \cite{ost89}:
\begin{equation}
L_{\beta} = Q(H) j_{B}/\alpha_{B}
\end{equation}
and
\begin{equation}
L_{\alpha} = Q(H) \frac{\alpha}{\beta}\frac{j_{B}}{\alpha_{B}}
\end{equation}
with  $j_{\beta}$, $\alpha_{B}$ and the Balmer ratio $\alpha/\beta$
that depend on the electronic temperature and, therefore, are different
for each metallicity. We use equations (6) and (8) from
\cite{fer80} for $j_{\beta}$ and $\alpha_{B}$. The ratio
$j_{\alpha}$/$j_{\beta}$ is given by \cite{ost89}.

We have measured the continuum of the stellar SEDs, $S_{\lambda}$, at
the corresponding wavelenghts, as well as the absorption line
luminosity which we have subtracted from the emission line, to
obtain the equivalent widths for the two lines:
\begin{equation}
EW(\AA) = \frac{[I_{emi}-I_{abs}](L_{\odot})}{S_{cont}(L_{\odot} \AA^{-1})}
\end{equation}
Table~\ref{balmer} gives the results of these calculations for a
cluster mass of 1 \Msun, a SAL2 IMF, solar metallicity and for the
first 5 Myr. The complete table is given in electronic format.  This
table shows for each age (column 1), the luminosity of the \hbeta\
line in emission (column 2), and in absorption (column 3); the value
of the continuum luminosity (stellar plus nebular) at 4860 \AA\
(column 4), the equivalent widths of \hbeta\ in emission (column 5),
in absorption (column 6) and total (column 7).  Columns 8 to 13 give
the same information for \halpha.These results are obtained by using
low resolution spectra. That means that the equivalent width in
absorption is not very precise and the errors are of the order of 5 to
10  per cent. The photoionization codes, like CLOUDY \citep*{fer98} computes
more precise values of emission lines and luminosities.

\begin{table*}
\caption{H$_{\beta}$ and H$_{\alpha}$ emission and absorption line
luminosity, continuum luminosity and equivalent widths, in emission,
in absorption and total.  Results are shown for the first 5 Myr
evolution of a cluster whose mass has been normalized to 1 \Msun\ with
SAL2 and \Zsun. The whole table is available in electronic format.}
\begin{tabular}{ccccccccccccc}
\hline
$\log{Age}$ & L$H{\beta}_{emi}$ & L$H{\beta}_{abs}$ &
$S_{cont,H\beta}$ & EW $H{\beta}_{emi}$ & EW $H{\beta}_{abs}$ &
EW $H{\beta}_{tot}$ & L$H{\alpha}_{emi}$ & L$H{\alpha}_{abs}$
& $S_{cont,H\alpha}$ & EW $H{\alpha}_{emi}$ & EW $H{\alpha}_{abs}$ &
EW$ H{\alpha}_{tot}$ \\ yr & \Lsun & \Lsun & \Lsun/\AA &
\AA & \AA & \AA & \Lsun & \Lsun & \Lsun/\AA & \AA &
\AA & \AA \\
\hline
 5.00 &  3.85 &  0.02 & 0.0080 &  479.26 &  2.66 &  476.60 &   11.49  & 0.25 & 0.0038 & 3002.76 & 64.78 & 2937.98 \\
 5.48 &  3.76 &  0.02 & 0.0082 &  459.45 &  2.54 &  456.91 &   11.22  & 0.24 & 0.0038 & 2914.03 & 61.70 & 2852.33 \\
 5.70 &  3.78 &  0.02 & 0.0083 &  454.63 &  2.53 &  452.10 &   11.29  & 0.24 & 0.0039 & 2893.54 & 61.70 & 2831.84 \\
 5.85 &  3.83 &  0.02 & 0.0085 &  452.14 &  2.52 &  449.63 &   11.42  & 0.24 & 0.0040 & 2883.57 & 61.69 & 2821.88 \\
 6.00 &  3.68 &  0.02 & 0.0088 &  417.89 &  2.54 &  415.35 &   10.97  & 0.25 & 0.0040 & 2730.14 & 61.75 & 2668.39 \\
 6.10 &  3.64 &  0.02 & 0.0091 &  399.45 &  2.53 &  396.92 &   10.85  & 0.25 & 0.0041 & 2643.12 & 61.78 & 2581.34 \\
 6.18 &  3.33 &  0.02 & 0.0094 &  354.16 &  2.57 &  351.59 &    9.94  & 0.25 & 0.0041 & 2415.59 & 61.87 & 2353.71 \\
 6.24 &  3.08 &  0.03 & 0.0098 &  314.93 &  2.59 &  312.34 &    9.20  & 0.26 & 0.0042 & 2208.67 & 61.96 & 2146.71 \\
 6.30 &  2.84 &  0.03 & 0.0105 &  271.55 &  2.60 &  268.95 &    8.46  & 0.27 & 0.0043 & 1961.56 & 62.04 & 1899.51 \\
 6.35 &  2.38 &  0.03 & 0.0110 &  215.76 &  2.63 &  213.13 &    7.11  & 0.27 & 0.0044 & 1619.65 & 62.18 & 1557.47 \\
 6.40 &  2.16 &  0.03 & 0.0120 &  181.14 &  2.62 &  178.52 &    6.46  & 0.29 & 0.0046 & 1394.58 & 62.24 & 1332.34 \\
 6.44 &  2.03 &  0.03 & 0.0131 &  155.79 &  2.65 &  153.13 &    6.07  & 0.31 & 0.0050 & 1221.75 & 62.34 & 1159.41 \\
 6.48 &  1.99 &  0.04 & 0.0140 &  142.07 &  2.63 &  139.45 &    5.92  & 0.33 & 0.0053 & 1124.17 & 61.96 & 1062.20 \\
 6.51 &  1.75 &  0.04 & 0.0143 &  122.29 &  2.74 &  119.55 &    5.21  & 0.33 & 0.0053 &  977.94 & 61.99 &  915.95 \\
 6.54 &  1.42 &  0.05 & 0.0143 &   99.56 &  3.19 &   96.37 &    4.24  & 0.33 & 0.0053 &  795.33 & 62.62 &  732.71 \\
 6.57 &  1.15 &  0.05 & 0.0139 &   82.70 &  3.53 &   79.17 &    3.43  & 0.33 & 0.0052 &  660.87 & 63.21 &  597.66 \\
 6.60 &  0.83 &  0.05 & 0.0141 &   58.61 &  3.84 &   54.77 &    2.47  & 0.33 & 0.0053 &  469.95 & 63.66 &  406.29 \\
 6.63 &  0.69 &  0.06 & 0.0135 &   50.75 &  4.15 &   46.60 &    2.05  & 0.32 & 0.0051 &  403.32 & 63.93 &  339.38 \\
 6.65 &  0.67 &  0.05 & 0.0118 &   56.17 &  4.17 &   52.00 &    1.98  & 0.29 & 0.0045 &  444.12 & 63.86 &  380.25 \\
 6.68 &  0.57 &  0.04 & 0.0098 &   57.98 &  3.84 &   54.14 &    1.70  & 0.23 & 0.0037 &  461.75 & 63.31 &  398.44 \\
 6.70 &  0.60 &  0.03 & 0.0093 &   63.87 &  3.53 &   60.34 &    1.78  & 0.22 & 0.0035 &  506.36 & 62.39 &  443.97 \\
\end{tabular}
\label{balmer}
\end{table*}

\subsection{Total Spectral Energy Distributions}
\begin{figure*}
\resizebox{\hsize}{!}{\includegraphics[angle=-90]{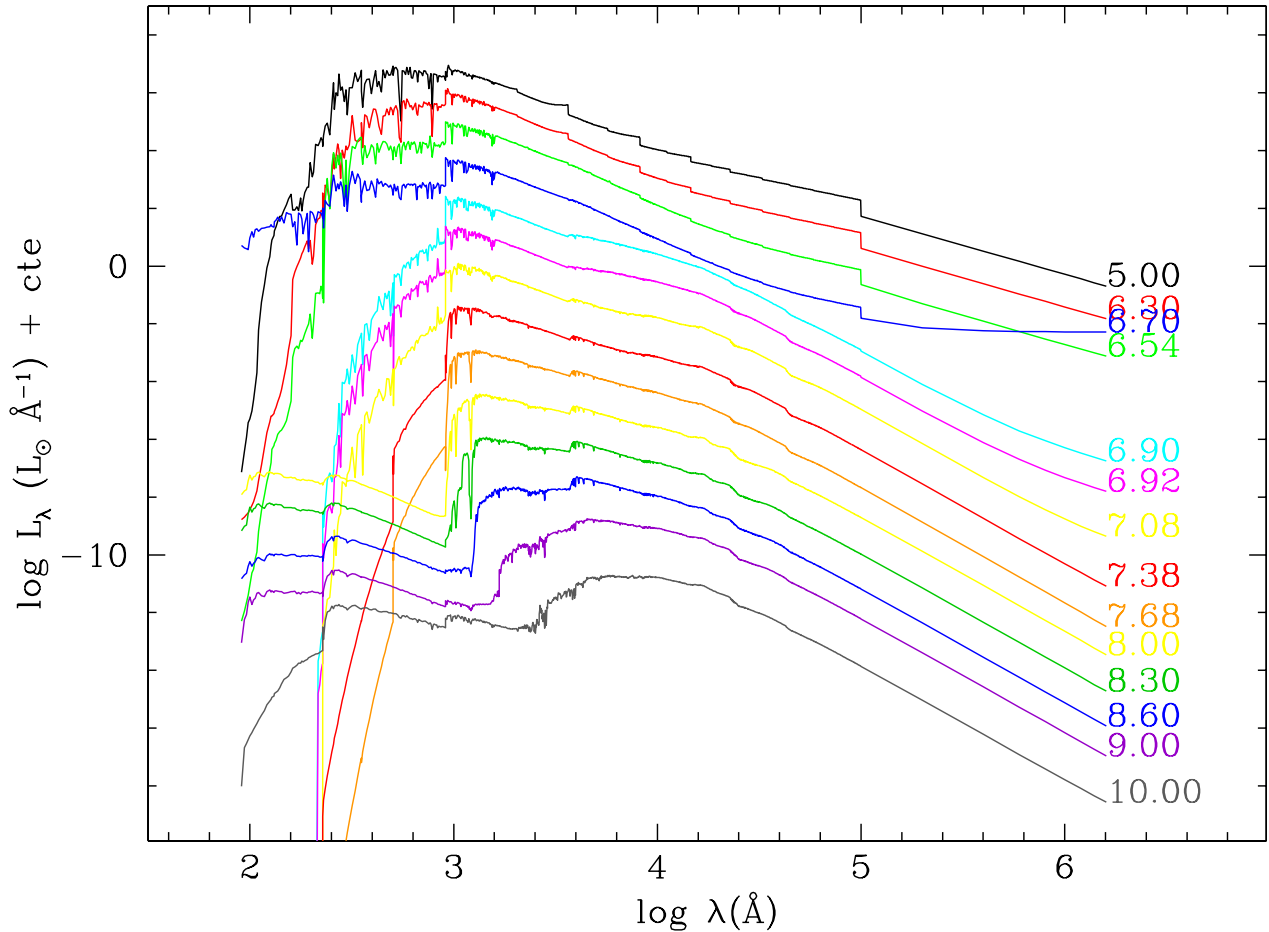}}
\caption{Spectral Energy Distributions for SAL2 and Z = \Zsun\
including both stellar and nebular contributions at the ages labelled
in logarithmic scale in the panel. Each age, starting at 0.1 Myr
($\log{\tau} = 5.00$), is shifted downward by one order of magnitude, for sake
of clarity.}
\label{spneb}
\end{figure*}

By adding the nebular contribution to the SSP SEDs we have obtained
the Total Spectral Energy Distributions. They are available in
electronic format in {http://www.fractal-es.com/SEDmod.htm} and in
compatible VO-table format in {http://esavo.esa.int/vospec/}, and can
be used in the VO (PopStar models).

Fig.~\ref{spneb} shows the total SEDs (arbitrary shifted for sake of
clarity) for the SAL2 IMF and Z = \Zsun\ and for some selected ages. The
nebular contribution is evident at the youngest ages ( $\log{\tau}
\leq 6.90$) and the ionization spectrum due to planetary nebula is
also evident for the intermediate and old age ones ($\log{\tau} \geq
8.60$).  Notice that the spectrum of the SSP at $\log{\tau} = 6.70$ is
dominated by WR stars.

\begin{figure}
\resizebox{\hsize}{!}{\includegraphics[angle=0]{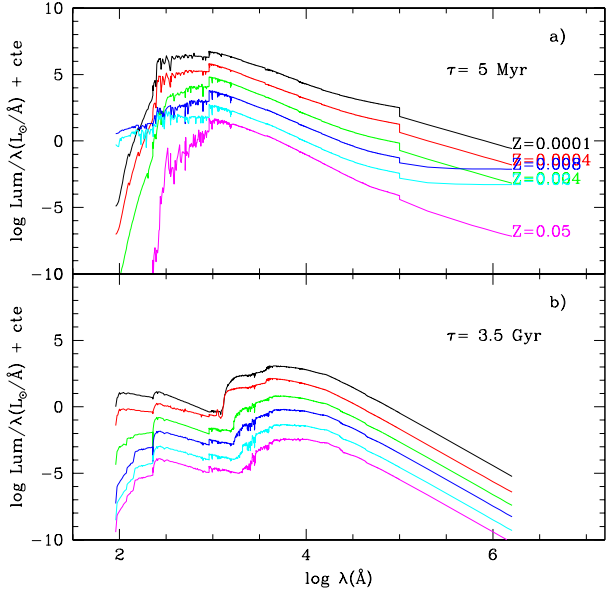}}
\caption{SEDs including both stellar and
nebular contributions for a SAL2 IMF and different metallicities, as
labelled. Panel a) for 5 Myr, and panel b) for 3.5 Gyr. Each SED
is shifted downward by one dex for sake of clarity.}
\label{spneb_zs}
\end{figure}

The dependence of the total SEDs on the metallicity is shown in
Fig.~\ref{spneb_zs} for two selected ages.  The largest differences
are found at ages younger than 10 Myr, where the different
contribution of the massive stars, due to different metallicity,
appears. For example at 10 Myr low metallicity SSPs are still capable
of producing a significant nebular continuum emission.  On the other
hand at these metallicities, there will not be WR single stars and,
consequently, the corresponding SSP will lack their hardening effects
at 5 Myr at shown in panel a) of Fig.~\ref{spneb_zs}.  At
intermediate and old ages --panel b of the same figure--
the differences at varying metallicity
mirror the effects of composition on the temperature of the turn-off
and of the RGB phase.  In this respect it is interesting to compare
with the age effects to appreciate the importance of the
age-metallicity degeneracy: we can see that similar variations in the
SED shape is found by changing either the age (Fig.~\ref{sps}) or the
metallicity (Fig.~\ref{spneb}b).

\section{Photometric evolution of stellar populations}

\subsection{Magnitudes and Colours}
In this section we provide  magnitudes and
colors computed for the Johnson-Cousins-Glass and for the SDSS systems.

Johnson-Cousins-Glass magnitudes UV1, UV2, U, B, V, R, I, J, H, K,
and L are computed using the definition suitable for photon counting devices
\citep*{gir02}:
\begin{equation}
m_{R_{\lambda}} = -2.5 \log{ \frac{\int_{\lambda_{1}}^{\lambda_{2}}
\lambda f_{\lambda} R_{\lambda}
d\lambda}{\int_{\lambda_{1}}^{\lambda_{2}} \lambda f^{0}_{\lambda}
R_{\lambda} d\lambda} } + m^{0}_{R_{\lambda}}
\end{equation}

where $R_{\lambda}$, the response function of the system,
$f_{\lambda}^{0}$, the reference spectrum (Vega in this system) and
its corresponding magnitudes, $m_{R_{\lambda}}^{0}$, completely define
the photometric system. For the reference spectrum we used a model
atmosphere corresponding to the Vega parameters, $\rm Teff = 9550$,
gravity $\log{g} = 4$ and $Z = 0.006$, interpolated on the
\cite{lcb97} spectral library.  Finally, absolute magnitudes were
computed assuming for Vega an absolute visual magnitude of $V = 0.58$
and all colors equal to zero.


Table~\ref{photometry} is as an example of the absolute magnitudes for
an SSP normalized to 1\Msun\, $Z = 0.02$ and a SAL2 IMF.  Age is in
column 1, column 2 and column 3 refer to the two ultraviolet HST
magnitudes UV1 and UV2 and the other columns provide the
U, B, V, R, J, H, K and L magnitudes.
\footnote{Magnitude B is given twice, the first one is used to compute
U-B while the second one is used for B-V.}.

\begin{figure*}
\resizebox{0.8\hsize}{!}{\includegraphics[angle = -90]{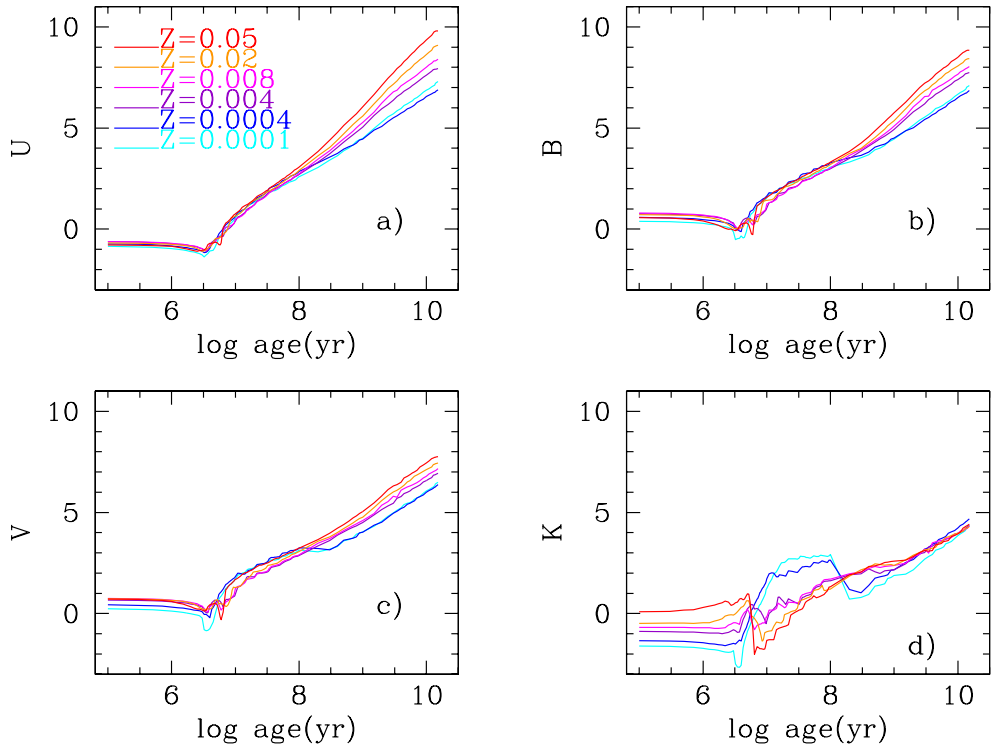}}
\caption{magnitudes evolution of magnitudes (Johnson system) for a stellar
population of 1 \Msun and SAL2 IMF: a) U, b) B, c) V and d) K. In each
panel the six metallicities are shown in different colours, as
labelled in panel a).}
\label{mag}
\end{figure*}

The evolution of U, B, V and K absolute magnitudes is shown in
Fig.~\ref{mag} for the case of the SAL2 IMF and for the six
metallicities considered here. Differences among metallicities are
more evident at old ages ($\tau > 1$ Gyr) except for K magnitude,
which shows a clear dependence on Z also for ($\tau < 1$ Gyr), due to
the important contribution of RSG first and AGB, post-AGB and red
giant at older ages.  At very young ages, $\tau \leq$ 6 Myr, an
important difference is due to the contribution of nebular emission.

\begin{figure*}
\resizebox{0.8\hsize}{!}{\includegraphics[angle=-90]{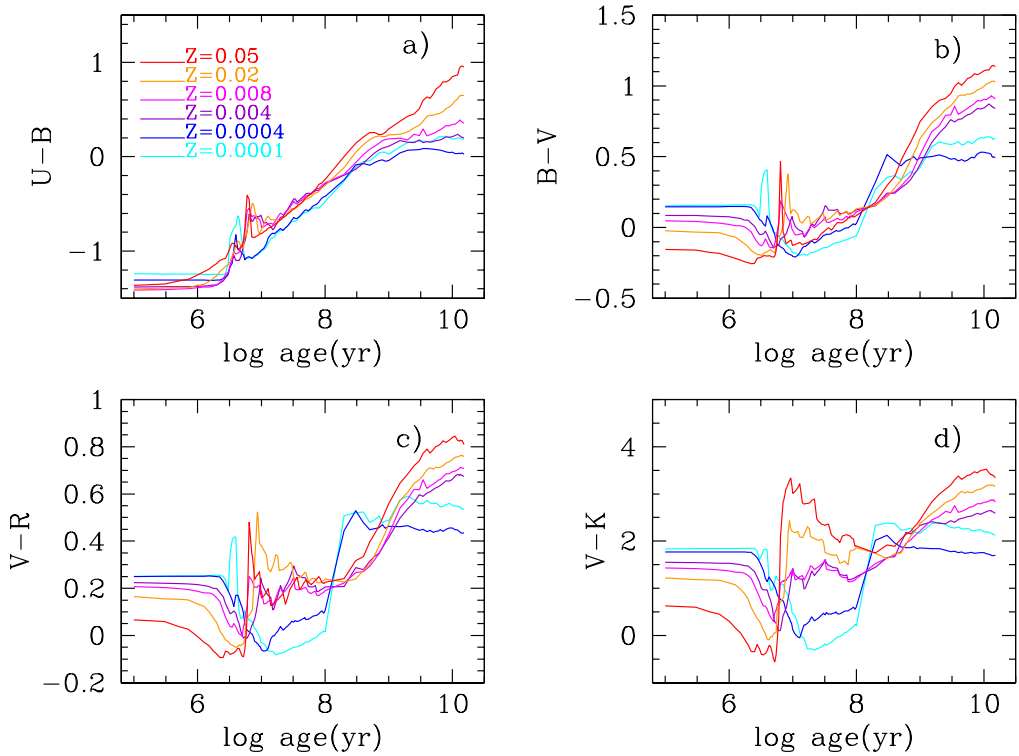}}
\caption{Clour evolution (Johnson system): a) U-B, b) B-V, c)
V-R and d) V-K. In each panel the six metallicities are shown in
different colours, as labelled in panel a).}
\label{col}
\end{figure*}

The evolution of the corresponding U-B, B-V, V-R and V-K colours is
shown in Fig.~\ref{col}.  Besides the well known effect of
age-metallicity degeneracy, other two important points appear from this
Figure.  Some colours do not increase monotonically with age. This is
particularly evident for the V-K colour and for solar and twice solar
metallicity.  The second point is the reddening of the colours (but
for the U-B) at ages less than a few Myr.  This effect, which is more
pronounced at low metallicity, is due to nebular continuum emission
being {\it relatively} more important at longer wavelengths.  For
example, the V-K colours, without nebular emission included, saturate
at V-K $\sim$-0.75 at ages below 3 Myr, for all metallicities.  Thus
the inclusion of nebular emission redden the colours of very young
populations significantly.  Though other effects may be at work at
these young ages, like dust attenuation, it is worth stressing that,
in presence of nebular emission, the colours of a very young metal
poor population are as red as those of an intermediate age metal rich
population.  These properties will be discussed in more details in the
forthcoming paper III where photometry including the emission lines,
contribution, computed in Paper II, will be calculated.

\begin{table*}
\caption{Example of the evolution of the magnitudes in the
Johnson-Cousins-Glass photometric
system, for a cluster mass normalized to 1 M$_{\odot}$, a Salpeter
IMF with mlow$ = 0.15 M_{\odot}$, mup$ =  100 M_{\odot}$, and Z$ = 0.02$.
The whole table is available in electronic format.}
\begin{tabular}{rrrrrrrrrrrrr}
\hline
$\log{Age}$ &  M$_{UV1}$ &  M$_{UV}$ &   M$_{U}$ &   M$_{B}$  &    M$_{B}$ &    M$_{V}$ &    M$_{R}$ &    M$_{I}$  &    M$_{J}$ &   M$_{H}$ &    M$_{K}$  &
M$_{L}$  \\
 yr &     &    & & &  & & & & & & \\
\hline
 5.00 & -4.267 & -1.973 & -0.680 &  0.723 &  0.714  & 0.737  & 0.572  & 0.562  & 0.382  & 0.073  &-0.483  &-1.448 \\
 5.48 & -4.311 & -2.005 & -0.692 &  0.698 &  0.689  & 0.722  & 0.567  & 0.561  & 0.391  & 0.088  &-0.464  &-1.424 \\
 5.70 & -4.333 & -2.026 & -0.709 &  0.679 &  0.670  & 0.706  & 0.553  & 0.549  & 0.381  & 0.079  &-0.471  &-1.431 \\
 5.85 & -4.354 & -2.047 & -0.727 &  0.660 &  0.650  & 0.688  & 0.537  & 0.533  & 0.366  & 0.065  &-0.485  &-1.444 \\
 6.00 & -4.419 & -2.106 & -0.755 &  0.608 &  0.599  & 0.657  & 0.523  & 0.529  & 0.377  & 0.086  &-0.454  &-1.406 \\
 6.10 & -4.463 & -2.150 & -0.784 &  0.567 &  0.557  & 0.626  & 0.502  & 0.513  & 0.371  & 0.085  &-0.449  &-1.397 \\
 6.18 & -4.520 & -2.203 & -0.799 &  0.521 &  0.510  & 0.604  & 0.504  & 0.528  & 0.412  & 0.142  &-0.375  &-1.310 \\
 6.24 & -4.567 & -2.256 & -0.824 &  0.469 &  0.458  & 0.573  & 0.495  & 0.531  & 0.441  & 0.187  &-0.314  &-1.235 \\
 6.30 & -4.627 & -2.327 & -0.870 &  0.390 &  0.378  & 0.514  & 0.461  & 0.513  & 0.456  & 0.223  &-0.256  &-1.157 \\
 6.35 & -4.691 & -2.392 & -0.898 &  0.320 &  0.307  & 0.470  & 0.450  & 0.523  & 0.516  & 0.317  &-0.123  &-0.991 \\
 6.40 & -4.737 & -2.463 & -0.959 &  0.228 &  0.215  & 0.394  & 0.396  & 0.482  & 0.510  & 0.336  &-0.072  &-0.909 \\
 6.44 & -4.775 & -2.532 & -1.033 &  0.128 &  0.116  & 0.305  & 0.323  & 0.419  & 0.474  & 0.322  &-0.056  &-0.864 \\
 6.48 & -4.799 & -2.578 & -1.091 &  0.053 &  0.041  & 0.234  & 0.260  & 0.362  & 0.430  & 0.291  &-0.069  &-0.857 \\
 6.51 & -4.743 & -2.552 & -1.085 &  0.026 &  0.014  & 0.212  & 0.249  & 0.359  & 0.448  & 0.330  &-0.001  &-0.750 \\
 6.54 & -4.567 & -2.419 & -1.019 &  0.034 &  0.022  & 0.207  & 0.248  & 0.361  & 0.468  & 0.378  & 0.095  &-0.592 \\
 6.57 & -4.421 & -2.295 & -0.938 &  0.072 &  0.061  & 0.235  & 0.280  & 0.392  & 0.509  & 0.439  & 0.197  &-0.429 \\
 6.60 & -4.284 & -2.186 & -0.884 &  0.061 &  0.050  & 0.221  & 0.272  & 0.379  & 0.516  & 0.478  & 0.302  &-0.217 \\
 6.63 & -4.137 & -2.047 & -0.782 &  0.111 &  0.101  & 0.263  & 0.308  & 0.403  & 0.533  & 0.501  & 0.352  &-0.112 \\
 6.65 & -4.045 & -1.934 & -0.645 &  0.257 &  0.247  & 0.405  & 0.445  & 0.534  & 0.648  & 0.603  & 0.436  &-0.061 \\
 6.68 & -3.953 & -1.831 & -0.494 &  0.452 &  0.441  & 0.611  & 0.655  & 0.750  & 0.866  & 0.812  & 0.636  & 0.119 \\
 6.70 & -3.925 & -1.808 & -0.457 &  0.509 &  0.498  & 0.672  & 0.709  & 0.799  & 0.901  & 0.831  & 0.638  & 0.096 \\
\hline
\end{tabular}
\label{photometry}
\end{table*}


Absolute magnitudes in the SDSS photometric system have been calculated
following  \citet*{gir04} and \citet{smith02}:
\begin{equation}
m_{AB} = -2.5\log{\frac{\int{\lambda f_{\lambda} R_{\lambda}d\lambda}}
{\int{\nu R_{\lambda}d\lambda}}}-48.60
\end{equation}
where  $R_{\lambda}$ are the response curves of the SDSS filters

\begin{figure*}
\resizebox{0.8\hsize}{!}{\includegraphics[angle=-90]{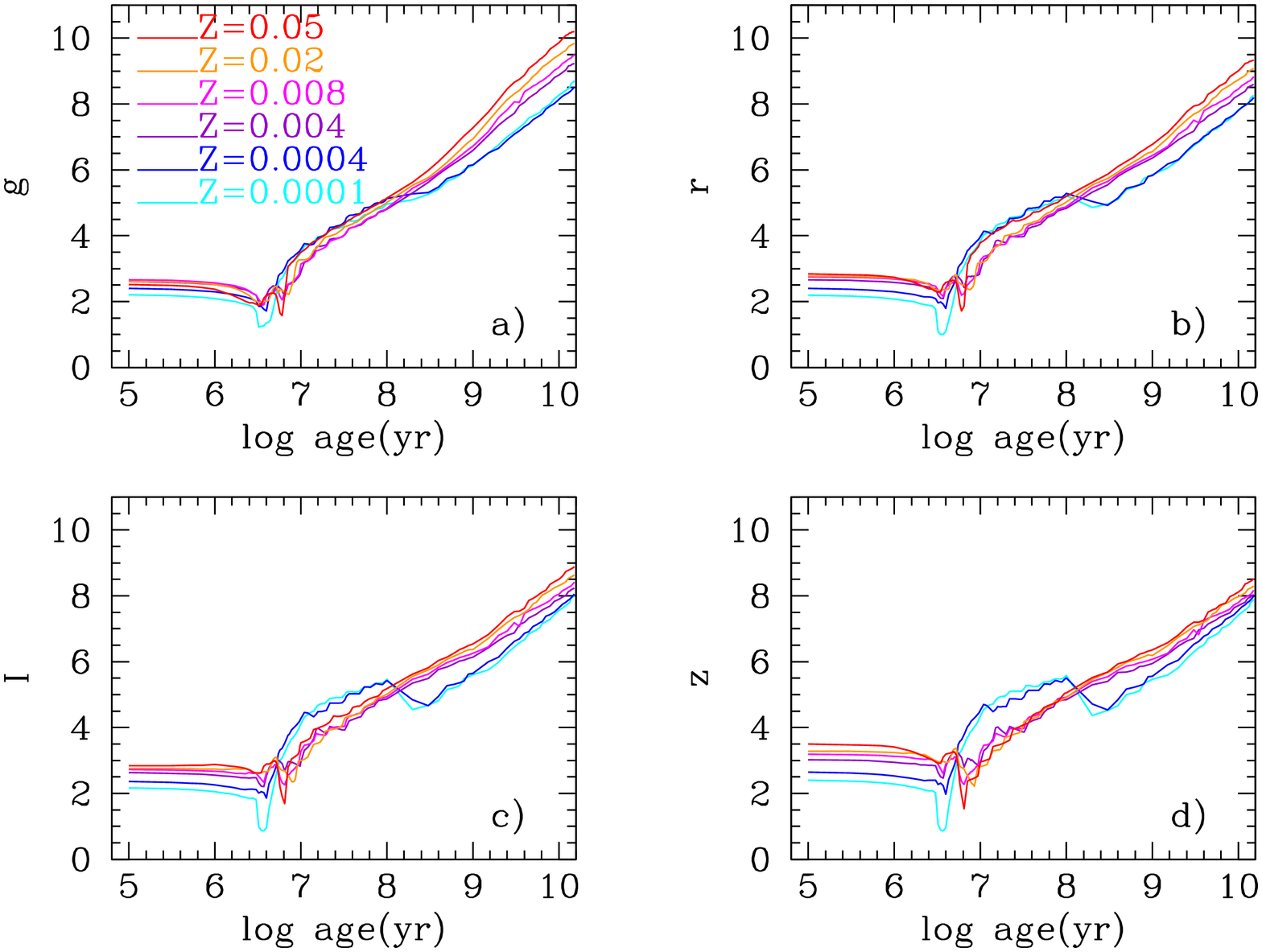}}
\caption{Evolution of SDSS magnitudes with age, for a stellar
population with mass normalized to 1 \Msun: a) g, b) r, c) I and d)
z. In each panel the six metallicities are shown in different colours, as
 labelled in panel a.}
\label{mag_sloan}
\end{figure*}

An example of absolute g, r, I and z magnitudes, in the first 5 Myr,
is provided in Table~\ref{photometry-SLOAN}. Their complete evolution
is shown in Fig.~\ref{mag_sloan}.

Magnitudes in these two photometric sytems for all IMFs, ages and
metallicities calculated  are available in electronic format in
{http://www.fractal-es.com/SEDmod.htm} and in compatible VO-tables.

\begin{figure}
\resizebox{\hsize}{!}{\includegraphics[angle=-90]{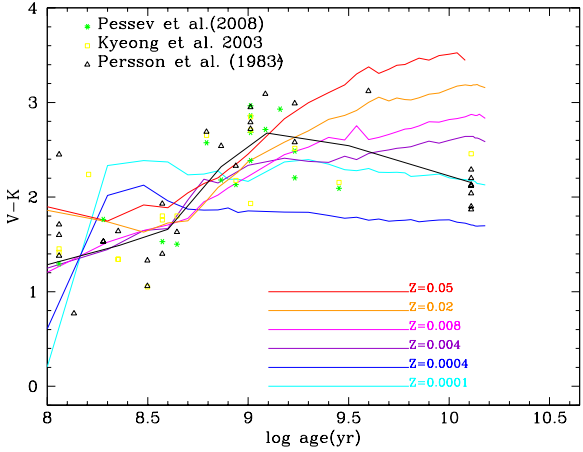}}
\caption{V-K colours of SSP compared with the observed extinction
corrected colours of LMC clusters.  Observed data are from authors
as labelled. In some cases the same cluster has observations from
different authors.}
\label{vktest}
\end{figure}

Before concluding this section, we show, in Fig.\ref{vktest}, the
comparison of our models against the observed integrated V-K colours
of LMC clusters from \citet{per83,kye03,goud06} and \citet{pes06}.
Cluster ages have been estimated by adopting the S-parameter
calibration by \citet*{girardi96}.  The black solid line is an average
within suitable age bins. Models of all the metallicities considered
here have been plotted.  Taking into account that the oldest LMC
clusters have very low metallicity, the intermediate age clusters have
Z $\sim$ 0.008 and the young clusters have about solar metallicity, we
may conclude that our models reproduce fairly well the observed colour
evolution of LMC clusters.

In the following we will compare our models, hereafter PopStar, with
those obtained by other authors and widely used in the literature.
Our comparison is limited to the models by \citet{stb99}, 
\citet[][ hereafter GALAXEV]{bc03}, \citet[][ hereafter
MAR05]{mar05}, \citet[][ hereafter GON05]{gon05}, and with our old
models by \citet[][hereafter MGV]{mgv00}.

\subsection{Comparison of results with Starburst99}

\begin{figure}
\resizebox{\hsize}{!}{\includegraphics[angle=0]{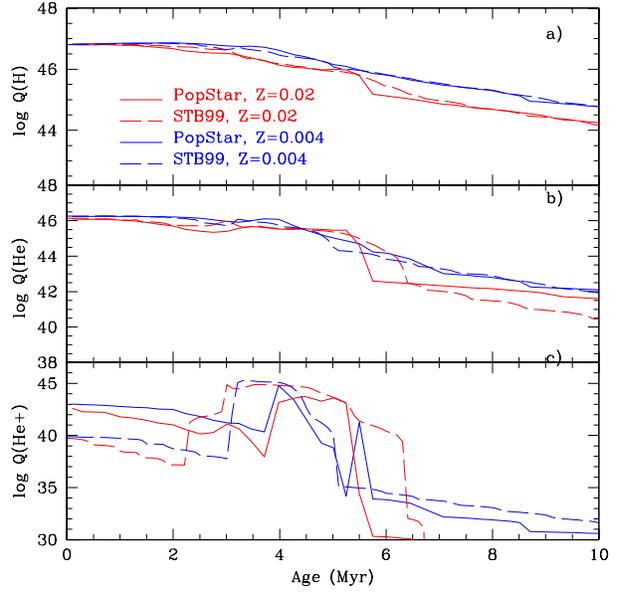}}
\caption{Comparison of the age evolution of the number of ionizing photons:
a) Q(H) b) Q(He), and c) Q(He+) of our models with the STB99 ones.
In each panel results for 2 metallicities are
represented  with a different line code as labelled in panel a).}
\label{qh_age_stb99}
\end{figure}

We begin our comparison with the models computed by Starburst99 code
recalling that, in its last version, Starburst99 includes the
\cite{snc02} atmosphere models for massive stars, as our models.
However for the young ages Starburst99 make use of the Geneva
isochrones.  Fig.~\ref{qh_age_stb99} shows the first 10 Myr evolution
of number of ionizing photons for H, He and He+ (top, medium and
bottom panel respectively).  In each panel, our model are depicted
with solid lines while, for STB99 we use long dashed lines. The blue
color lines correspond to a low metallicity (Z = 0.004) case while, the
red ones are for the solar metallicity (Z = 0.02). Differences between
STB99 and this work are not very large for Q(H) and Q(He). They are
more important for Q(He+) in the case of solar metallicity where our
models show a more extended WR phase.

\begin{table}
\caption{SDSS magnitudes for the first 5 Myr of a cluster of 1
M$_{\odot}$, with a SAL2 IMF and Z$ = 0.02$.  The whole table is available
in electronic format.}
\begin{tabular}{rrrrr}
\hline
$\log{\tau}$ &  g &  r &   I &   z  \\
 yr         &    &    &     &      \\
\hline
 5.00 &   2.615 &  2.779 &  2.765 &  3.287  \\
 5.48 &   2.595 &  2.770 &  2.767 &  3.282  \\
 5.70 &   2.577 &  2.755 &  2.755 &  3.269  \\
 5.85 &   2.558 &  2.738 &  2.739 &  3.253  \\
 6.00 &   2.515 &  2.715 &  2.738 &  3.240  \\
 6.10 &   2.478 &  2.689 &  2.723 &  3.220  \\
 6.18 &   2.441 &  2.679 &  2.743 &  3.223  \\
 6.24 &   2.397 &  2.658 &  2.750 &  3.215  \\
 6.30 &   2.326 &  2.611 &  2.737 &  3.183  \\
 6.35 &   2.266 &  2.582 &  2.755 &  3.173  \\
 6.40 &   2.179 &  2.516 &  2.720 &  3.118  \\
 6.44 &   2.083 &  2.434 &  2.662 &  3.044  \\
 6.48 &   2.009 &  2.366 &  2.607 &  2.980  \\
 6.51 &   1.982 &  2.346 &  2.609 &  2.967  \\
 6.54 &   1.983 &  2.337 &  2.622 &  2.954  \\
 6.57 &   2.017 &  2.363 &  2.662 &  2.972  \\
 6.60 &   2.003 &  2.351 &  2.663 &  2.941  \\
 6.63 &   2.050 &  2.388 &  2.698 &  2.953  \\
 6.65 &   2.195 &  2.526 &  2.830 &  3.084  \\
 6.68 &   2.395 &  2.737 &  3.040 &  3.307  \\
 6.70 &   2.452 &  2.798 &  3.086 &  3.359  \\
 \end{tabular}
\label{photometry-SLOAN}
\end{table}

\begin{figure}
\resizebox{\hsize}{!}{\includegraphics[angle=0]{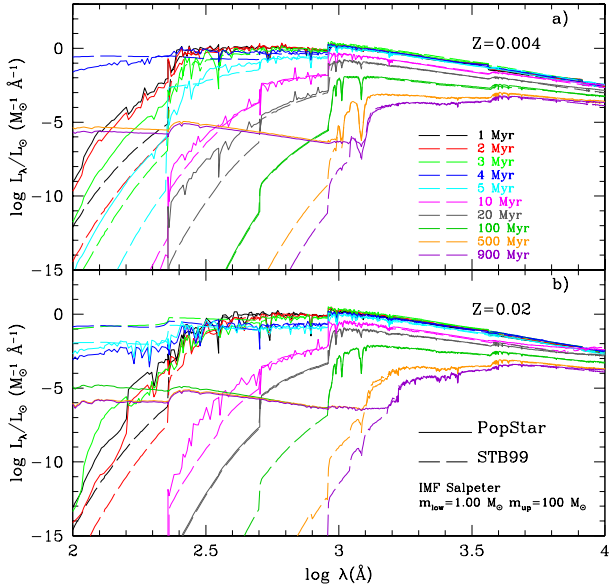}}
\caption{Comparison between SEDs obtained with PopStar and the
corresponding ones from STB99 for different ages. Each age is plotted
with a color. Solid lines are PopStar models while dashed lines are
STB99 ones.}
\label{sp_comp_stb99}
\end{figure}

In Fig.~\ref{sp_comp_stb99} we compare our SEDs with those obtained
with STB99, for some ages between 1 and 900 Myr.  Panel a) shows the case
Z$ = $0.004 while panel b) shows referes to Z$ = $0.02. In both cases an
STB IMF has been assumed and a normalization to a total initial mass
of 1 M$_\odot$.  Some differences appears at young ages, mainly due to
isochrones differences as mentioned above.  Remarkable differences
appear at ages between 100-900 Myr when, in our isochrones, Planetary
Nebulae appear. The nature of this discrepancy is certainly the higher
fuel of our Post AGB stars, due to the different mass loss rate
prescriptions adopted in our revised isochrones, besided the use of
the model spectra from \cite{rau03}.

\begin{figure}
\resizebox{\hsize}{!}{\includegraphics[angle=0]{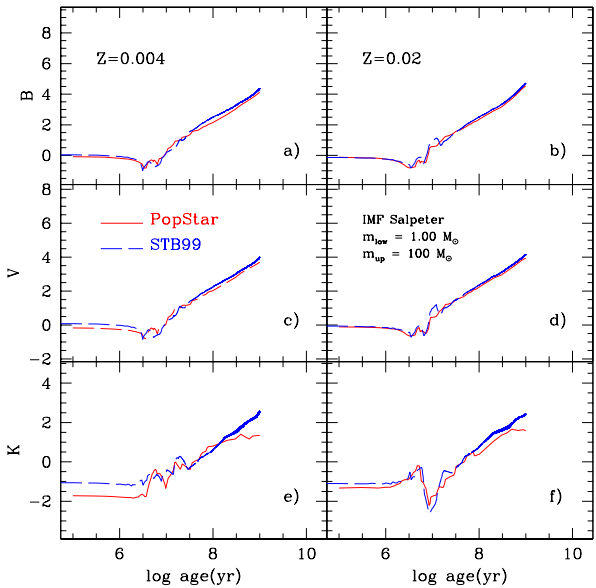}}
\caption{Magnitude evolution in B,V and K bands obtained in this work
compared with the corresponding from STB99.  Solid red lines are
PopStar models while the dashed blue lines are STB99 ones.}
\label{mag_comp_stb}
\end{figure}

Fig.~\ref{mag_comp_stb} shows the evolution of B, V and K magnitudes
in the top, medium and bottom panels, respectively. In the left hand
side panels we show the results for Z$ = $0.004, while those for
Z$ = $0.02 are shown in the right hand side panels.  Our PopStar results
are represented by red solid lines while for STB99 we use blue dashed
lines.  Notice that STB99 models are calculated only until 1 Gyr.
Magnitudes B and V are practically identical in both cases. However,
the K band shows slight differences with brighter (redder) stellar
population for the youngest ages in our models compared with the STB99
ones, mostly for Z$ = $0.004, at low metallicity. The same occurs for
ages $\tau > 100$ Myr: STB99 models are less bright than ours in K
band.

\begin{figure}
\resizebox{\hsize}{!}{\includegraphics[angle=0]{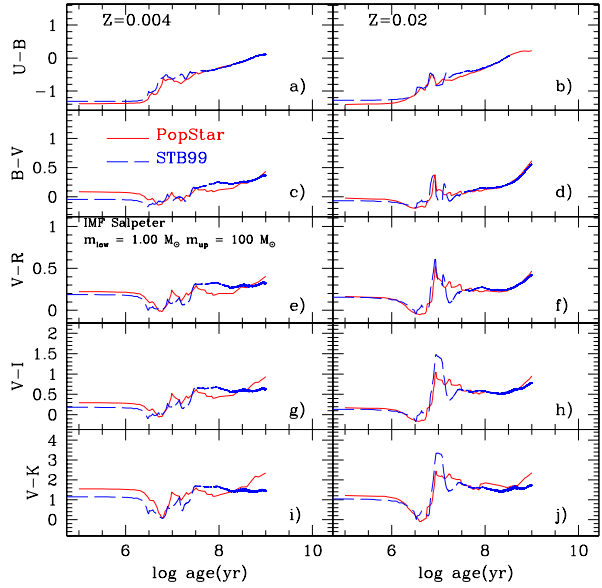}}
\caption{Color evolution U-B, B-V, V-R, V-I and V-K obtained in this
work compared with the ones in STB99.  Solid red lines are PopStar
results while the dashed blue lines are STB99 ones.}
\label{col_comp_stb}
\end{figure}
The comparison of the U-B, B-V, V-R, V-I and V-K colours is shown in
Fig.~\ref{col_comp_stb}, adopting the same notation for the models.
The agreement is good. However, at solar metallicity there are some
differences in the RSG colours (especially V-K) while, at low
metallicity, there are differences in the AGB phase (likely caused by
different mass loss rates) and at very young ages (likely caused by
differences in nebular emission continuum contribution).

\subsection{Comparison of results with GALAXEV models}

\begin{figure}
\resizebox{\hsize}{!}{\includegraphics[angle=0]{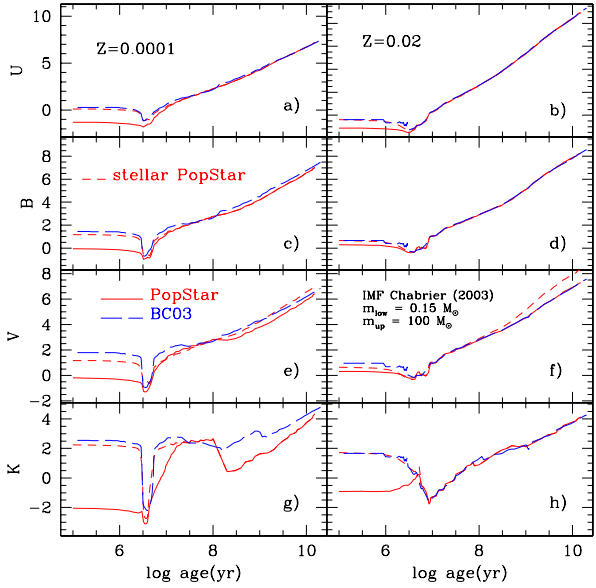}}
\caption{Magnitude evolution in U, B, V and K with the age obtained in
this work compared with the corresponding ones from GALAXEV. Solid red
lines are PopStar models while the dashed blue lines are those models
from GALAXEV. Dashed red lines are the results obtained for the stellar
PopStar models without the nebular contribution.}
\label{mag_comp_bc03}
\end{figure}

\begin{figure}
\resizebox{\hsize}{!}{\includegraphics[angle=0]{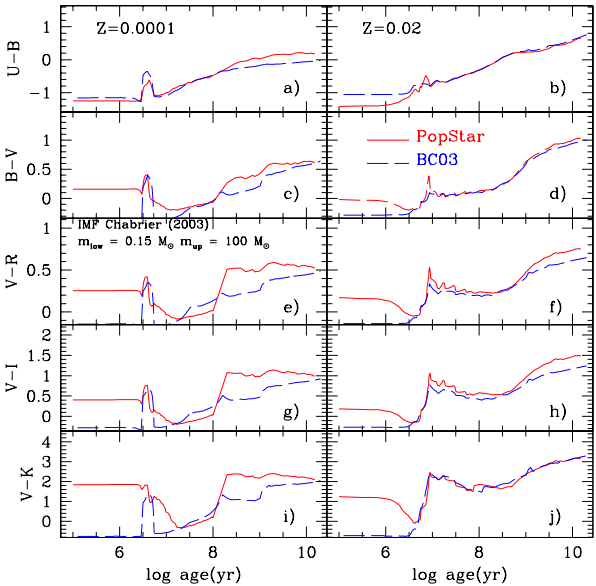}}
\caption{Color evolution: U-B, B-V, V-R, V-I and V-K colors obtained
for PopStar compared with the corresponding ones from GALAXEV. The
same coding line than previous figures has been used.}
\label{col_comp_bc03}
\end{figure}

Here we compare our results with GALAXEV models \citep{bc03} that are
particularly well suited for old populations.  Fig~\ref{mag_comp_bc03}
shows the comparison of the U, B, V and K magnitudes while,
Fig.~\ref{col_comp_bc03}, shows the comparison of of U-B, B-V, V-R,
V-I and V-K colours. For all models we have assumed a Chabrier IMF
with solid red lines indicating our PopStar models and dashed blue
lines depicting the corresponding GALAXEV models.  Left hand panels
show the evolution for Z$ = $0.0001 while, right hand panels, refer to
Z$ = $0.02.  At young ages our colours are significantly redder than
those of BC03, mainly because BC03 do not include the effects of
nebular continuum. To show this we plot in Fig.~\ref{mag_comp_bc03} as
short-dashed red lines the results obtained for the PopStar stellar spectra
without the continuum nebular contribution. The results are very similar
to those ones from BC03. The difference PopStar-BC03 increases at 
decreasing metallicity
because the strength of the nebular emission increases at decreasing
metallicity.  As already said, including this contribution at young
ages, produces a reddening of the V-K colours from two to three
magnitudes, depending on the initial metallicity.  This contribution
is thus relevant and should be included in the interpretation of the
spectral evolution of star forming galaxies.  At intermediate and old
ages the differences in magnitudes and in colours are small, in
particular for the case of solar metallicity. At low metallicity the
difference become larger, especially in the V-K. The origin of this
difference is most likely a different algorithm to describe the mass
loss rates.  This can be seen also from Fig.~\ref{sp_comp}, where we
perform a comparison of the spectral evolution of a solar metallicity
SSP, predicted by different authors.  We notice that, contrary to the
other models, a significant Post AGB phase is already present after
100 Myr in our models.  From the same figure we may also notice the
effects of the inclusion of the new spectra of Planetary Nebulae.
Both effects combine to produce a P-AGB phase with a lower far UV
luminosity.

\subsection{Comparison of results with Maraston models}

\begin{figure}
\resizebox{\hsize}{!}{\includegraphics[angle=0]{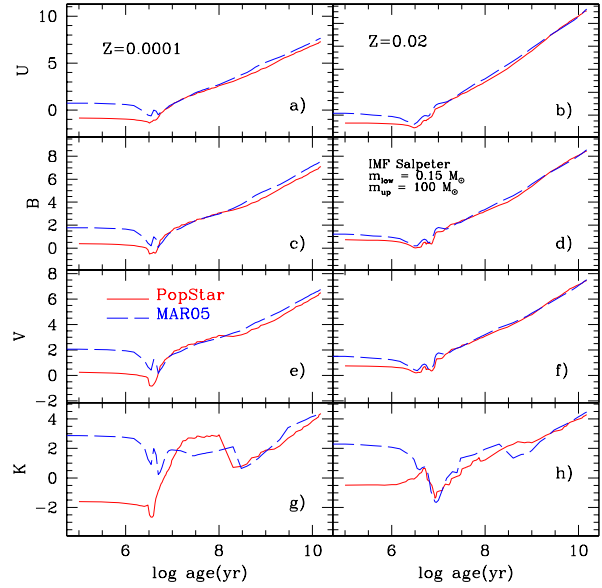}}
\caption{Magnitude evolution in bands U, B, V and K for PopStar,
compared to the corresponding ones from MAR05 with the same coding
line than previous figures as labelled in the plot.}
\label{mag_comp_mar}
\end{figure}

\begin{figure}
\resizebox{\hsize}{!}{\includegraphics[angle=0]{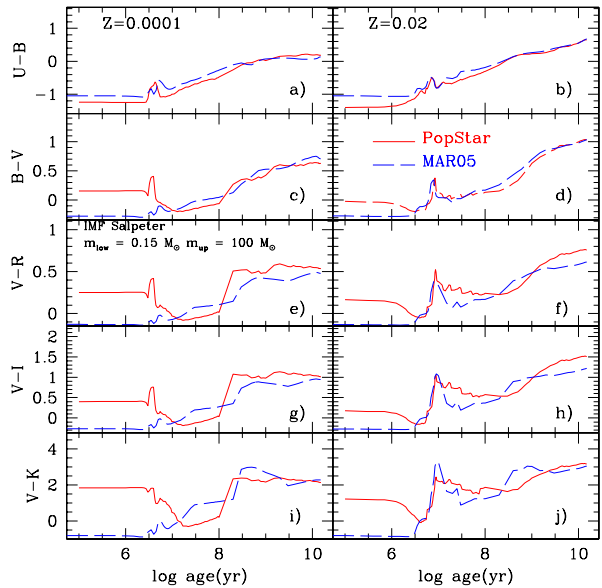}}
\caption{Color evolution showing U-B, B-V, V-R, V-I and V-K with age
obtained in this work compared with the corresponding ones from MAR05
with the same coding line than previous figures.}
\label{col_comp_mar}
\end{figure}

We have also compared our PopStar results with MAR05 models.  The
evolution of selcted magnitudes and colours are shown in
Fig.~\ref{mag_comp_mar} and Fig.~\ref{col_comp_mar}, respectively.  We
have used the same symbols and line coding used in the previous
figures.

Since MAR05 does not include nebular emission, a large difference is
present at very young ages, as before. There are also differences at older ages
which may reach a magnitude in the V-K colour. The differences with
respect to MAR05 models are in general larger than those with respect
to BC03 models



\begin{figure*}
\resizebox{\hsize}{!}{\includegraphics[angle=-90]{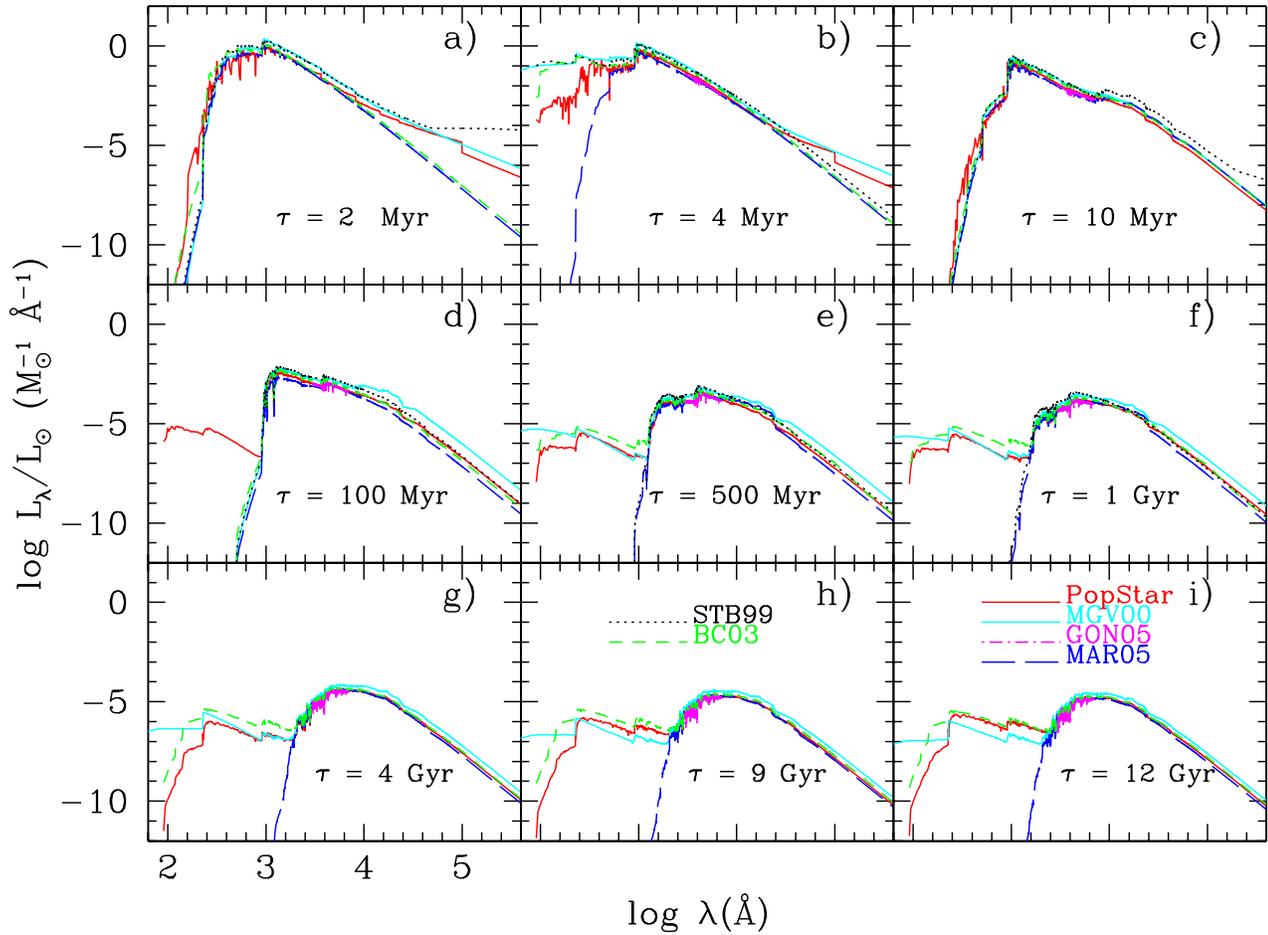}}
\caption{SEDs comparison. Our models, PoPStar, are plotted together
with those from MGV00, GON05, MAR05, STB99 and GALAXEV(BC03). Each model is
represented by a different type of line as coding as labelled.
Different ages sampling the whole evolution are plotted.}
\label{sp_comp}
\end{figure*}

Finally, Fig.~\ref{sp_comp} shows the comparison of the spectral
evolution computed with PopStar for a SSP with solar
metallicity and a SAL2 IMF  with other codes existing in
literature, \citet{stb99} named as STB99, \citet{bc03} as BC03, \citet{mar05}
as MAR05, \citet{gon05} as GON05, and with our old models, \citet{mgv00}
as MGV00, at some selected ages, as given in each panel.

From this figure it appears that MAR05 models lack both the hard
ionizing flux of the Wolf Rayet stars as well as that of Post AGB
stars. On the other hand STB99 models lack the the Post-AGB
phase. GON05 models were calculated with higher resolution but within
a shorter wavelength range (from 3000 to 7000 \AA) than the other
models. The only models that include the hard ionizing flux of WR
stars and that of P-AGB stars are those by BC03 and our, old and new,
ones.

Our new models are similar to our old MGV models. However the
different procedure of temperature assignation and the different
spectral library adopted, in the WR phase, result in a significantly
less hard ionizing spectra This effects is also present in the
Post-AGB phase. 
 The nebular continuum is now more accurate because
it includes not only the hydrogen but also the helium
contribution.  Furthermore, in the common wavelength range, our new
models compare very well also with GON05 models.


From Fig.~\ref{sp_comp} we see that our models reproduce very well the
behavior of the STB99 models at young ages.  Moreover PopStar models
follow very well the evolution of the GALAXEV and MAR05 models at old
ages.  Thus since STB99 models were specifically calibrated for
starburst galaxies while GALAXEV and MAR05 were mainly calibrated for
intermediate/old ages, we conclude that PopStar models can be safely
used over the whole age range of stellar populations.


\section{Conclusions}

We have computed new synthesis models that follow the evolution of
simple stellar populations from very young to very old ages.  To this
purpose we have combined the homogeneous and well calibrated set of
Padova 94 stellar evolution models, with the most recent physics input
concerning stellar atmospheres and, where needed, nebular continuum
emission.  Besides spectra over a wide wavelength range, these models
provide a large quantity of additional information, useful for the
analysis of the properties of stellar populations (see Appendix A).

From the comparison of our models with previous evolutionary synthesis
models, we have shown that our models are well calibrated and can be
safely applied over the whole age range of stellar populations.  This
is particularly important because we have shown that not all existing
models include all the relevant evolutionary phases of stellar
populations

For this reason our models are particularly suited for the study of
galaxies where a combination of stellar populations of different ages
and metallicities coexist.


Our new models, the resulting SEDs and HR diagrams are available in
the Virtual Observatory (Models in http://esavo.esa.int/vospec/ and
the PopStar web-page at http://www.fractal-es.com/SEDmod.htm).

\section{Acknowledgments}

This work has been partially supported by DGICYT grant
AYA2007--67965-C03-03.  and by the Spanish MEC under the
Consolider-Ingenio 2010 Program grant CSD2006-00070: First Science
with the GTC (http://www.iac.es/consolider-ingenio-gtc/).
A.B. acknowledges contract ASI-INAF I/016/07/0.

\appendix
\label{appA}
\section{Products available}

The resulting SEDs  are available in the VO (Models in
http://esavo.esa.int/vospec/ and both SEDs and  HR diagrams
are also available at http://www.fractal-es.com/SEDmod.htm).

The products already available in the PoPStar web-page described in
this paper are: (2) SSP models for different IMF parameters, ages and
metallicities; (1) H-R Diagrams for different IMF parameters, ages and
metallicities; (3) Results tables with SSPs's properties related to
the cluster ionization power: numbers of O, B, WN, WC, WO, SN, PN,
ionizing photons (Q(H), Q(He), Q(He$^{+}$, Q(O)), etc. Other available
products in the web-page are the emission line spectra of the ionized
gas resulting from the ionizing clusters (described in Paper II) and
the photometrical parameters, in which the colours have been computed
taking into account the contribution of the emission line spectra for
the youngest ages. The photometrical results are presented and
discussed in Paper III.

\begin{enumerate}

\item H-R diagrams
\begin{itemize}
\item M/\Msun-HR = Initial mass (\Msun)
\item Teff-HR = Effective Temperature (K) as computed in the isochrone
\item logg-HR = Logarithm of the gravity (cms$^{-2}$) from the
isochrone
\item log(L/\Lsun)-HR = Logarithm of the Luminosity (in \Lsun), from
the isochrone
\item Teff-Md = Effective Temperature (K) as computed of the assigned
model
\item logg-Md = Logarithm of the gravity (cms$^{-2}$)
\item log(L/\Lsun)-Md = Logarithm of the Luminosity (in \Lsun),
\item log(Mdot)= Logarithm of the mass loss (in \Msun yr$^{-1}$)
\item Err(Teff)= Error associated to the Teff (K) assignation
\item Err(logL/\Lsun) = Error associated to the Log(L/\Lsun)
assignation
\item WR = Parameter with the following meaning: 1 (WN star), 2 (WC
star), 3 (evolved hot star, e.g. PN), 0 �normal� stars (all the
stars non-included in the previous categories)
\item nstar-sal1 = Number of stars in that mass interval (weights)
with IMF-1 normalized to 1 \Msun\ cluster
\item nstar-sal2 = Number of stars in that mass interval (weights)
with IMF-2 normalized to 1 \Msun\ cluster
\item nstar-fer = Number of stars in that mass interval (weights) with
IMF-3 normalized to 1 \Msun\ cluster
\item nstar-kro = Number of stars in that mass interval (weights) with
IMF-4 normalized to 1 \Msun\ cluster
\item nstar-cha = Number of stars in that mass interval (weights) with
IMF-5 normalized to 1 \Msun\ cluster
\end{itemize}

\item Spectral Energy Distributions: We provide the spectra (SEDs)
normalized to 1 \Msun\ cluster.  Each file has four columns with the
following contents:
\begin{itemize}
\item $\lambda$ (\AA) = Wavelength
\item Lstar$_{\lambda}$ / \Lsun (\AA$^{-1}$) = Star continuum
\item Lneb$_{\lambda}$ / \Lsun (\AA$^{-1}$) = Nebular continuum
\item Ltot$_{\lambda}$ / \Lsun (\AA$^{-1}$) = Total (star+nebular)
continuum
\end{itemize}

Spectra can be chosen by IMF type, age and metallicity. All the
spectra have been normalized to a total mass of 1 \Msun.  To convert
the luminosities in ergs$^{-1}$ you need to multiply by the solar
luminosity, \Lsun $= 3.82 \times 10^{33}$ ergs$^{-1}$. Spectra can be
managed with VOSpec

\item Parameters relevant to the ionization cluster power: we provide
3 tables with the following information (all the numbers have been
normalized to 1 \Msun).
\begin{itemize}
\item logage = Logarithm of the age (in yr)
\item N$_{tot}$ = Total number of stars at the time of cluster
formation
\item N$_{20}$ = Total number of stars with M $leq 20$ \Msun
\item N$_{OB}$ = Total number of O, B stars present at the current age
\item N$_{WN}$ = Total number of WN stars present at the current age
\item N$_{WC}$ = Total number of WC stars present at the current age
\item N$_{RSG}$= Total number of RSG stars present at the current age
\item N$_{SN}$ = Total number of SNs Ib+II exploded in each age step
\item N$_{PN}$ = Total number of PN present at the current age
\item E$_{winds}$ = Total Mechanical Energy deposited by winds at the
current age
\item E$_{SNs}$ = Total Mechanical Energy deposited by SNs in each age step
\item R$_{c}$ = Cluster radius at the current age (from mechanical
energy deposition due to winds+SNs)
\item QH = Total number of HI ionizing photons, Q(H),
$\lambda <$ 912.4 \AA\ in s$^{-1}$
\item QHe = Total number of HeI ionizing photons, Q(He),
$\lambda <$ 504.6 \AA\ in s$^{-1}$
\item QO = Total number of OI ionizing photons, Q(O),
$\lambda <$ 353.3 \AA\ in s$^{-1}$
\item QHe$^{+}$ = Total number of HeII ionizing photons,
Q(He$^{+}$), $\lambda <$ 228.0 \AA\ in s$^{-1}$
\end{itemize}
\end{enumerate}

\end{document}